\documentclass[11pt,a4paper]{article}
\usepackage{jcappub}
\usepackage{amssymb,amsmath,amsfonts}
\usepackage{bm}
\usepackage{amsfonts}
\usepackage{latexsym}
\usepackage{graphicx}
\usepackage{amsmath}
\usepackage{palatino}
\usepackage{mathpazo}
\usepackage{textcomp}
\usepackage{subfigure}
\usepackage{float}
\usepackage{booktabs}
\usepackage{mathtools}
\usepackage{dcolumn}
\usepackage{ragged2e}

\usepackage{hyperref}
\hypersetup{
    colorlinks=true,
    linkcolor=red,
    citecolor=blue,
    filecolor=magenta,      
    urlcolor=cyan,
    pdftitle={Can $f(R)$ gravity isotropize a pre-bounce contracting universe?},
    pdfpagemode=FullScreen} 
\usepackage{amsmath}
\usepackage{xcolor}
\usepackage{orcidlink}

\title{\boldmath A dynamical systems formulation for inhomogeneous LRS-II spacetimes}

\author[a,b]{Saikat Chakraborty\orcidlink{0000-0002-5472-304X}}
\author[c,b,d]{Peter K.S. Dunsby\orcidlink{0000-0002-6271-9585}}
\author[e]{Rituparno Goswami\orcidlink{0000-0002-5355-3288}}
\author[b,f]{Amare Abebe\orcidlink{0000-0001-5475-2919}}
\affiliation[a]{The Institute for Fundamental Study ``The Tah Poe Academia Institute", Naresuan University, Phitsanulok 65000, Thailand} 
\affiliation[b]{Centre for Space Research, North-West University, Potchefstroom 2520, South Africa}
\affiliation[c]{Department of Mathematics and Applied Mathematics, Cosmology and Gravity Group,  University of Cape Town, Rondebosch, 7701, Cape Town, South Africa} 
\affiliation[d]{South African Astronomical Observatory, Observatory 7925, Cape Town, South Africa} 
\affiliation[e]{Astrophysics and Cosmology Research Unit,
School of Mathematics, Statistics and Computer Science,
University of KwaZulu-Natal,
Private Bag X54001, Durban 4000, South Africa}
\affiliation[f]{National Institute for Theoretical and Computational Sciences (NITheCS), 3201 Stellenbosch, South Africa}
\emailAdd{saikat.ch@nu.ac.th}
\emailAdd{peter.dunsby@uct.ac.za}
\emailAdd{goswami@ukzn.ac.za}
\emailAdd{amare.abebe@nithecs.ac.za}

\abstract{We present a dynamical system formulation for inhomogeneous LRS-II spacetimes using the covariant 1+1+2 decomposition approach. Our approach describes the LRS-II dynamics from the point of view of a comoving observer. Promoting the covariant radial derivatives of the covariant dynamical quantities to new dynamical variables and utilizing the commutation relation between the covariant temporal and radial derivatives, we were able to construct an autonomous system of first-order ordinary differential equations along with some purely algebraic constraints. Using our dynamical system formulation we found several interesting features in the LRS-II phase space with dust, one of them being that the homogeneous solutions constitute an invariant submanifold. For the particular case of LTB, we were also able to recover the previously known result that an expanding LTB tends to Milne in the absence of a cosmological constant, providing a potential validation of our formalism.}

\begin{document}

\maketitle


\section{Introduction}

One of the key pillars of the standard cosmological model is the assumption of the cosmological principle (sometimes called the Copernican principle). According to this principle, averaged over a cosmologically relevant scale, say a few percent of the particle horizon, it is a good enough approximation to assume that matter in the universe is distributed in a smooth spatially homogeneous and isotropic manner. It is further assumed that the spacetime geometrical features trace that of the matter distribution. Together, these two assumptions give rise to the modelling of the cosmological spacetime with the Friedmann-Lema\'itre-Robertson-Walker (FLRW) metric. The standard model of cosmology, the so-called $\Lambda$CDM model is based on the FLRW spacetime geometry. Small-scale inhomogeneities are treated as cosmological perturbations \cite{Tsagas:2007yx}.

Nevertheless, as is well known, the $\Lambda$CDM model suffers from the cosmological constant problem \cite{Weinberg:1988cp}, $H_0$-problem \cite{DiValentino:2020zio} and $\sigma_8$ problem \cite{DiValentino:2020vvd}. These problems fuelled the search for alternative models to describe the observable universe. Most attempt to find alternative models to describe a dynamical dark energy, as opposed to a ``constant'' dark energy given by the cosmological constant. These alternative models are constructed by adding one or more additional dynamical degrees of freedom to the known theory, either by adding additional matter components (canonical or non-canonical fields, non-ideal fluids or combinations thereof \cite{Copeland:2006wr,Amendola_Tsujikawa_2010}) or altering the theory of gravity altogether (scalar-tensor theories dominating this direction \cite{papantonopoulos2014modifications,CANTATA:2021ktz}). Of course, there are clear problems associated with both approaches. If one adds additional matter components, one needs to worry about their explanation from the particle physics side. On the other hand, modified gravity theories are usually highly constrained by local gravity tests and many a time are plagued by various stability issues. For an extensive review of the dynamical analysis of such alternative models of dark energy, the reader is referred to the beautiful review by Bahamonde et. al. \cite{Bahamonde:2017ize}.

An interesting alternative line of explaining the apparent acceleration of the universe without the introduction of any new dynamical degree of freedom to the theory is based on relaxing the cosmological principle, in particular the assumption of homogeneity. Since a full non-perturbative treatment of generic inhomogeneous spacetime dynamics is very difficult, the toy model of spherically symmetric inhomogeneous spacetimes, namely the Lema\'itre-Tolman-Bondi (LTB) model was considered. One class of works have studied LTB models with a purely formal notion of acceleration \cite{Nambu:2005zn,Moffat:2005ii,Apostolopoulos:2006eg,Mansouri:2005rf}. Another class of works focused on whether it is possible to explain without assuming any notion of an acceleration the observed relationship between the luminosity distance of redshift \cite{Vanderveld:2006rb,Alnes:2005rw,Garfinkle:2006sb,Iguchi:2001sq}, the very relationship which gave rise to the notion of an accelerated universe under the assumption of homogeneity. Of course, even these models come with their fair share of observational constraints \cite{Zumalacarregui:2012pq,Marra:2011ct,Perivolaropoulos:2014lua}. Nonetheless, two explicit theorems were proved by Mustapha, Hellaby and Ellis in \cite{Mustapha:1997fjz} which make it clear that homogeneity cannot be proven without either a fully determinate theory
of source evolution or availability of distance measures that are independent of the source evolution. This theorem makes it worthwhile to further study inhomogeneous cosmology. Inhomogeneous LTB models are also the starting point of studies on realistic gravitational collapse \cite{Joshi:2000fk,Joshi:2011zm,Malafarina:2016let}.

Dynamical system analysis has proved to be an invaluable tool for studying the dynamics of homogeneous spacetimes. Since in standard cosmology we mostly deal with homogeneous spacetimes, the dynamical system approach in cosmology is an invaluable companion to theoretical cosmologists \cite{Wainwright_Ellis_1997,Coley:1999uh}. For homogeneous dynamical spacetimes, this formalism recasts the gravitational field equations as a system of autonomous first-order ordinary differential equations along with some algebraic constraints. The gravitational dynamics are then recast as a flow in a phase space spanned by suitably defined dimensionless dynamical variables (which are usually normalized by the expansion parameter of a timeline congruence). When a self-gravitating system is inhomogeneous, as in an LTB cosmology or LTB collapse model, the techniques of the dynamical system cannot be applied in a straightforward manner. The reason is that now the governing equations of the system are partial differential equations.

That is not to say that such attempts have not been made. One such approach that could be found in the literature utilizes the so-called orthonormal frame formalism \cite{vanElst:1996dr}. The issue with this approach is that both the evolution and constraint equations are partial differential equations, involving temporal and spatial derivatives respectively. Although some techniques from dynamical systems theory could be applied, the analysis becomes mathematically intricate \cite{vanElst:2001xm,Uggla:2003fp,Lim:2003ut}. Another approach that is taken by Sussman introduces the so-called quasi-local variables \cite{Sussman:2007ku,Sussman:2010zp}. Sussman's approach does indeed give rise to an effective system of ordinary differential equations with algebraic constraints, but the quasilocal variables by themselves do not directly correspond to the actual covariant geometric and thermodynamic quantities relevant for a comoving observer.

In this work, we present a new dynamical system formulation for a class of inhomogeneous spacetimes with a local rotational symmetry, namely the LRS-II. The motivation to consider this class is that it contains the spherically symmetric solutions e.g. LTB that are of physical interest. Our formulation is based on the 1+1+2 covariant decomposition approach \cite{Clarkson:2002jz,Clarkson:2007yp}, which itself is built up on the 1+3 covariant decomposition approach \cite{Ehlers:1961xww,Ellis:1998ct}  \footnote{That 1+1+2 covariant approach can potentially be helpful in constructing a dynamical system formulation for LRS-II spacetimes is hinted in \cite{Ganguly:2014qia,Cruz:2017ecg}. However, the authors considered only LRS-II vacuum or electrovacuum solutions for which case it reduces to either static or homogeneous spacetime  by virtue of the Birkhoff's theorem.}. To proceed with this approach one needs to first specify a timeline unit 4-vector $u^{\mu}$ and a spacelike unit 4-vector $e^{\mu}$, which we choose to be along the worldline of a comoving observer and the direction of the axis of local rotational symmetry respectively. In the covariant decomposition approaches, all the dynamical quantities have a very clear and coordinate-independent physical interpretation. Physically, our approach involves going into the frame of a comoving observer and following the dynamics from his/her point of view. The effect of spatial inhomogeneity is that not all the comoving observers would experience the same evolutionary phase (characterized by a fixed point), simultaneously, a point we also try to elaborate on in the main text. The novelty of our approach is that we could formulate an autonomous system with only time-dependent evolution equations and algebraic constraints and our dynamical variables are also directly related to covariant geometrical and thermodynamic quantities relevant to a comoving observer.

The paper is organized as follows. In section \ref{sec: lrs2_dyn} we present an exposition to LRS-II spacetimes using the 1+1+2 covariant formalism. We dedicate a complete subsection discussing various intricacies presented by the inhomogeneity of the system. In section \ref{sec: lrs2_ps} we construct an autonomous dynamical system for LRS-II using expansion normalized dimensionless dynamical variables. We dedicate an entire subsection to discussing the interpretation of the dynamical structure, something which we deem important here. Sections \ref{sec: inv_sub}, \ref{sec: fin_fp} and \ref{sec: inf_fp} present the invariant submanifolds, finite fixed points and asymptotic fixed points in the LRS-II phase space. In section \ref{sec: ltb} we specialize to the spherically symmetric LTB model and discuss its phase space. We conclude in section \ref{sec: conclusion} summarizing the important and novel aspects of our work with an outlook of possible further research along this line. Finally, appendix \ref{sec: comp_Rpos} is devoted to exploring a possible different route to constructing an autonomous system for cases with positive 3-curvatures, which proves to be not very fruitful.


\section{Some basics about covariant splitting}

The well-known 1+3 covariant approach to cosmology pioneered by Ehlers \cite{Ehlers:1961xww} and Ellis \cite{Ellis:1998ct} starts by splitting all the covariant geometric and thermodynamic quantities into a part that lies along a preferred timelike unit 4-vector $u^{\mu}$ ($u^{\mu}u_{\mu}=-1$) and one perpendicular to it. The specific choice of a 4-vector essentially chooses a ``frame''. The metric tensor $g_{\mu\nu}$ can be decomposed as
\begin{equation}\label{13}
    g_{\mu\nu} = U_{\mu\nu} + h_{\mu\nu},
\end{equation}
where the tensor $U_{\mu\nu}=-u_{\mu}u_{\nu}$ projects along $u^{\mu}$ and the tensor $h_{\mu\nu}$ projects on the 3-surface orthogonal to $u^{\mu}$ ($h_{\mu\nu}u^{\nu}=0$). A 4-scalar quantity cannot be further decomposed. A 4-vector $V^{\mu}$ can be irreducibly decomposed into a scalar and a 3-vector using Eq.\eqref{13}:
\begin{eqnarray}
V^{\mu} = - \mathcal{V}u^{\mu} + \mathcal{V}^{\mu}\;,
\end{eqnarray}
where the scalar $\mathcal{V}=V^{\mu}u_{\mu}$ is the projection of $V^{\mu}$ along the 4-vector $u^{\mu}$ and the 3-vector $\mathcal{V}^{\mu}=h^{\mu}_{\alpha}V^{\alpha}$ is the projection of $V^{\mu}$ on the 3-surface orthogonal to $u^{\mu}$. In the language of covariant 1+3 splitting, we can say $\mathcal{V}^{\mu}$ is a 3-vector because it completely lies on the 3-surface orthogonal to $V^{\mu}$ ($\mathcal{V}^{\mu}u_{\mu}=0$). 

A 4-tensor can be irreducibly decomposed into a trace part, which is a scalar, and a tracefree part. The tracefree part can be further decomposed using Eq.\eqref{13}. For example, a tracefree 2nd rank 4-tensor $T_{\mu\nu}$ can be irreducibly decomposed into a scalar, 3-vectors and a projected tracefree 3-tensor using Eq.\eqref{13} as follows:
\begin{eqnarray}
T_{\mu\nu} = \mathcal{T}\left(u_{\mu}u_{\nu} + \frac{1}{3}h_{\mu\nu}\right) + \mathcal{T}_{1\mu}u_{\nu} + \mathcal{T}_{2\nu}u_{\mu} + \mathcal{T}_{\mu\nu},
\end{eqnarray}
where we have defined
\begin{subequations}
\begin{eqnarray}
&& \mathcal{T}\equiv T^{\alpha\beta}u_{\alpha}u_{\beta} = T^{\alpha\beta}h_{\alpha\beta}\;, \\
&& \mathcal{T}_{1\mu} \equiv -T^{\alpha\beta}h_{\alpha\mu}u_{\beta}, \quad \mathcal{T}_{2\mu}\equiv -T^{\alpha\beta}u_{\alpha}h_{\beta\mu}\;, \\
&& \mathcal{T}_{\mu\nu}\equiv \left(h_{\mu\alpha}h_{\nu\beta} - \frac{1}{3}h_{\mu\nu}h_{\alpha\beta}\right)T^{\alpha\beta}\;.
\end{eqnarray}
\end{subequations}
In the language of the 1+3 covariant splitting, we can say $\mathcal{T}_{1\mu},\,\mathcal{T}_{2\mu}$ and $\mathcal{T}_{\mu\nu}$ are 3-vectors and 3-tensors, respectively, because they lie completely on the 3-surface orthogonal to $u^{\mu}$ ($\mathcal{T}_{1\mu}u^{\mu}=0=\mathcal{T}_{2\mu}u^{\mu}$, $\mathcal{T}_{\mu\nu}u^{\mu}=0$). Also, note that the 3-tensor $\mathcal{T}_{\mu\nu}$ is tracefree ($h_{\mu\nu}\mathcal{T}^{\mu\nu}=0$). 

The covariant derivative of a 4-vector and a 2nd rank 4-tensor can be decomposed respectively as
\begin{subequations}
    \begin{eqnarray}
        && \nabla_{\mu}V^{\alpha} = - \dot{V}^{\alpha}u_{\mu} + D_{\mu}V^{\alpha} - u^{\alpha}u_{\gamma}h^{\nu}_{\,\,\,\mu}\nabla_{\nu}V^{\gamma}\,,
        \\
        && \nabla_{\mu}T^{\alpha}_{\,\,\,\beta} = -\dot{T}^{\alpha}_{\,\,\,\beta}u_{\mu} + D_{\mu}T^{\alpha}_{\,\,\,\beta} + h^{\nu}_{\,\,\,\mu}u^{\alpha}u^{\sigma}u_{\gamma}u_{\beta}\nabla_{\nu}T^{\gamma}_{\,\,\,\sigma} - (u^{\alpha}u_{\gamma}h^{\sigma}_{\,\,\,\beta} + u^{\sigma}u_{\beta}h^{\alpha}_{\,\,\,\gamma})h^{\nu}_{\,\,\,\mu}\nabla_{\nu}T^{\gamma}_{\,\,\,\sigma}\;,\notag
        \\
    \end{eqnarray}
\end{subequations}
where we have defined the covariant derivative along the timelike 4-vector $u^{\mu}$ and the fully orthogonally projected covariant derivative on the 3-surface respectively as
\begin{subequations}
    \begin{eqnarray}
    \dot{V}^{\alpha} = u^{\mu}\nabla_{\mu}V^{\alpha}\,, &\quad& 
    \dot{T}^{\alpha}_{\,\,\,\beta} = u^{\mu}\nabla_{\mu}T^{\alpha}_{\,\,\,\beta}\,,
    \\
    D_{\mu}V^{\alpha} = h^{\nu}_{\,\,\,\mu}h^{\alpha}_{\,\,\,\gamma}\nabla_{\nu}V^{\gamma}\,, &\quad& D_{\mu}T^{\alpha}_{\,\,\,\beta} = h^{\nu}_{\,\,\,\mu}h^{\alpha}_{\,\,\,\gamma}h^{\sigma}_{\,\,\,\beta}\nabla_{\nu}T^{\gamma}_{\,\,\,\sigma}\,.
    \end{eqnarray}
\end{subequations}
It is straightforward to generalize the above definitions for higher-rank 4-tensors.

The 1+1+2 covariant approach \cite{Clarkson:2002jz,Clarkson:2007yp} builds on the 1+3 approach by further splitting the three spatial degrees of freedom into a part that lies along a preferred spacelike 4-vector $e^\mu$ and one perpendicular to $e^\mu$ (referred to as the ``sheet''). The vector $e^{\mu}$ is taken to be normalized and it lies completely on the 3-surface orthogonal to $u^{\mu}$
\begin{equation}
    e^{\mu}e_{\mu}=1, \quad u^{\mu}e_{\mu}=0.
\end{equation}
In the language of 1+3 covariant splitting, we can say $e^{\mu}$ is a 3-vector.
With respect to $e^{\mu}$, the 3-tensor $h_{\mu\nu}$ is further decomposed as
\begin{equation}\label{112}
    h_{\mu\nu} = E_{\mu\nu} + N_{\mu\nu},
\end{equation}
where the 3-tensor $E_{\mu\nu}=e_{\mu}e_{\nu}$ projects along $e^{\mu}$ and the 3-tensor $N_{\mu\nu}$ projects on the 2-sheet orthogonal to both $u^{\mu}$ and $e^{\mu}$ ($N_{\mu\nu}u^{\mu}=0=N_{\mu\nu}e^{\mu}$). Using Eq. \eqref{112} a 3-vector $\mathcal{V}^{\mu}$ can be irreducibly decomposed into a scalar and a 2-vector
\begin{eqnarray}
\mathcal{V}^{\mu} = \mathbb{V}e^{\mu} + \mathbb{V}^{\mu}\;,
\end{eqnarray}
where the scalar $\mathbb{V}=\mathcal{V}^{\mu}e_{\mu}$ is the projection of the 3-vector $\mathcal{V}^{\mu}$ along the 3-vector $e^{\mu}$ and the 2-vector $\mathbb{V}^{\mu}=N^{\mu\alpha}\mathcal{V}_{\alpha}$ is the projection of the 3-vector $\mathcal{V}^{\mu}$ on the 2-dimensional sheet orthogonal to $e^{\mu}$. We say $\mathbb{V}^{\mu}$ is a 2-vector because it completely lies on the 2-sheet orthogonal to both $u^{\mu}$ and $e^{\mu}$ ($\mathbb{V}^{\mu}u_{\mu}=0=\mathbb{V}^{\nu}e_{\nu}$). 

Similarly, a tracefree 3-tensor can be further decomposed using Eq. \eqref{112}. For example, a tracefree 2nd rank 3-tensor $\mathcal{T}_{\mu\nu}$ can be irreducibly decomposed into a scalar, 2-vectors and a projected tracefree 2-tensor using Eq. \eqref{112} as follows:
\begin{eqnarray}
\mathcal{T}_{\mu\nu} = \mathbb{T}\left(e_{\mu}e_{\nu} - \frac{1}{2}N_{\mu\nu}\right) + \mathbb{T}_{1\mu}e_{\nu} + \mathbb{T}_{2\nu}e_{\mu} + \mathbb{T}_{\mu\nu},
\end{eqnarray}
where we have defined
\begin{subequations}
\begin{eqnarray}
&& \mathbb{T} \equiv \mathcal{T}^{\alpha\beta}e_{\alpha}e_{\beta} = - \mathcal{T}^{\alpha\beta}N_{\alpha\beta}, \\
&& \mathbb{T}_{1\mu} \equiv \mathcal{T}^{\alpha\beta}N_{\alpha\mu}e_{\beta}, \quad \mathbb{T}_{2\mu} \equiv \mathcal{T}^{\alpha\beta}e_{\alpha}N_{\beta\mu} \\
&& \mathbb{T}_{\mu\nu}\equiv \left(N_{\mu\alpha}N_{\nu\beta} - \frac{1}{2}N_{\mu\nu}N_{\alpha\beta}\right)\mathcal{T}^{\alpha\beta}
\end{eqnarray}
\end{subequations}
$\mathbb{T}_{1\mu},\,\mathbb{T}_{2\mu}$ and $\mathbb{T}_{\mu\nu}$ are 2-vector and 2-tensor, respectively, because they lie completely on the 2-sheet orthogonal to $u^{\mu}$ and $e^{\mu}$ ($\mathbb{T}_{1\mu}u^{\mu}=\mathbb{T}_{1\mu}e^{\mu}=0=\mathbb{T}_{2\mu}e^{\mu}=\mathbb{T}_{2\mu}u^{\mu}$, $\mathbb{T}_{\mu\nu}u^{\mu}=0=\mathbb{T}_{\mu\nu}e^{\mu}$). Also, note that the 2-tensor $\mathbb{T}_{\mu\nu}$ is tracefree ($N_{\mu\nu}\mathbb{T}^{\mu\nu}=0$). 

The fully orthogonally projected covariant derivative of a 3-vector and a 2nd rank 3-tensor can be decomposed respectively as
\begin{subequations}
    \begin{eqnarray}
        && D_{\mu}\mathcal{V}^{\alpha} = \hat{\mathcal{V}}^{\alpha}e_{\mu} + d_{\mu}\mathcal{V}^{\alpha} + e^{\alpha}e_{\gamma}N^{\nu}_{\,\,\,\mu}D_{\nu}V^{\gamma}\,,
        \\
        && D_{\mu}\mathcal{T}^{\alpha}_{\,\,\,\beta} = \hat{\mathcal{T}}^{\alpha}_{\,\,\,\beta}e_{\mu} + d_{\mu}\mathcal{T}^{\alpha}_{\,\,\,\beta} + N^{\nu}_{\,\,\,\mu}e^{\alpha}e^{\sigma}e_{\gamma}e_{\beta}D_{\nu}\mathcal{T}^{\gamma}_{\,\,\,\sigma} + (e^{\alpha}e_{\gamma}N^{\sigma}_{\,\,\,\beta} + e^{\sigma}e_{\beta}N^{\alpha}_{\,\,\,\gamma})N^{\nu}_{\,\,\,\mu}D_{\nu}\mathcal{T}^{\gamma}_{\,\,\,\sigma}\,,\notag
        \\
    \end{eqnarray}
\end{subequations}
where we have defined the covariant derivative along the spacelike 3-vector $e^{\mu}$ and the fully orthogonally projected covariant derivative on the 2-sheet respectively as:
\begin{subequations}
    \begin{eqnarray}
    \hat{\mathcal{V}}^{\alpha} = e^{\mu}D_{\mu}\mathcal{V}^{\alpha}\,, &\quad& 
    \hat{\mathcal{T}}^{\alpha}_{\,\,\,\beta} = e^{\mu}D_{\mu}\mathcal{T}^{\alpha}_{\,\,\,\beta}\,,
    \\
    d_{\mu}\mathcal{V}^{\alpha} = N^{\nu}_{\,\,\,\mu}N^{\alpha}_{\,\,\,\gamma}D_{\nu}\mathcal{V}^{\gamma}\,, &\quad& d_{\mu}\mathcal{T}^{\alpha}_{\,\,\,\beta} = N^{\nu}_{\,\,\,\mu}N^{\alpha}_{\,\,\,\gamma}N^{\sigma}_{\,\,\,\beta}D_{\nu}\mathcal{T}^{\gamma}_{\,\,\,\sigma}\,.
    \end{eqnarray}
\end{subequations}
It is straightforward to generalize the above definitions for higher-rank 3-tensors.

Eqs. \eqref{13} and \eqref{112} can be written in a combined form as
\begin{equation}
    g_{\mu\nu} = U_{\mu\nu} + E_{\mu\nu} + N_{\mu\nu} = - u_{\mu}u_{\nu} + e_{\mu}e_{\nu} + N_{\mu\nu},
\end{equation}
which can be used to decompose \emph{all} the covariant physical quantities into scalars, 2-vectors and 2-tensors, with the later two lying completely on the 2-dimensional sheet orthogonal to $u^{\mu},\,e^{\mu}$. The beautiful aspect of the 1+1+2 decomposition formulation is that, for spacetimes with a local rotational symmetry, if the spatial 2-vector $e^{\mu}$ is taken at each spatial point to be along the axis of the local rotational symmetry, all the 2-vectors and 2-tensors vanish, leaving us only with a set of scalars that characterize the dynamics. 

When the metric is diagonal and the temporal and the radial coordinates are affine parameters along $u^{\mu}$ and $e^{\mu}$ respectively, then the explicit form of these vectors are
\begin{equation}
    u^{\mu}=\left(\frac{1}{\sqrt{-g_{00}}},0,0,0\right)\,, \quad e^{\mu}=\left(0,\frac{1}{\sqrt{g_{rr}}},0,0\right)\,.
\end{equation}
The covariant derivatives along these two vectors for a scalar quantity are related to its ordinary time and radial derivatives as follows:
\begin{subequations}
    \begin{eqnarray}
        && \dot{\psi} = u^{\mu}\nabla_{\mu}\psi = u^{\mu}\partial_{\mu}\psi = \frac{1}{\sqrt{-g_{00}}}\partial_{t}\psi\,,\label{dot_derivative}
        \\
        && \hat{\psi} = e^{\mu}D_{\mu}\psi = e^{\mu}h^{\nu}_{\mu}\nabla_{\nu}\psi = e^{\mu}\partial_{\mu}\psi = \frac{1}{\sqrt{g_{rr}}}\partial_{r}\psi\,.\label{hat_derivative}
    \end{eqnarray}
\end{subequations}

The above formulation is completely covariant, i.e. coordinate-independent. When a particular coordinate system is chosen and the metric is specified, different covariant quantities can be calculated explicitly in terms of the metric functions. 


\section{Dynamics of LRS-II spacetime with a perfect fluid}
\label{sec: lrs2_dyn}

We consider the LRS-II class of spacetimes that admits spherically symmetric solutions and is rotation-free. The most general LRS-II metric can be written as \cite{Stewart:1967tz}
\begin{equation}\label{LRS-II}
    ds^2 = - \frac{1}{A^2(t,r)}dt^2 + B^2(t,r)dr^2 + C^2(t,r)[dy^2 + D_k^2(y)dz^2]\,,
\end{equation}
where $t,\,r$ are affine parameters along the preferred timeline 4-vector $u^{\mu}$ and the preferred spacelike 4-vector $e^{\mu}$ and $k=(-1,0,1)$ denotes open, flat and closed geometry of the 2-sheets respectively. The function $D_k(y)$ is
\begin{equation}
    D_k(y) = 
    \begin{cases}
    \sin{y}, \qquad k=+1 \\
    y, \qquad k=0 \\
    \sinh{y}, \qquad k=-1
    \end{cases}.
\end{equation}
Over the last few years, the covariant 1+1+2 splitting approach, originally introduced by Clarkson in Ref.\cite{Clarkson:2002jz} to study perturbations around the Schwarzschild background, has been used extensively in the study of LRS-II spacetimes, both in General Relativity and in $f(R)$ gravity \cite{Betschart:2004uu, Nzioki:2016wmj, Nzioki:2010nj, Nzioki:2009av}. In general, for LRS-II spacetimes, there is a set of evolution equations containing only `dot'-derivatives, a set of propagation equations containing only `hat'-derivatives and a set of mixed evolution-propagation equations containing both hat and dot derivatives. The amazing simplicity offered by confining our attention only to perfect fluids is that there we can get rid of the mixed equations. The evolution and the propagation equations can be completely decoupled. 

The complete set of 1+1+2 covariant equations for LRS-II spacetimes in the presence of a generic fluid can be found in \cite{Clarkson:2007yp}. Here we only give the definitions of the key quantities related to this problem, together with their evolution and constraint equations. 

The set of non-vanishing covariant geometric and thermodynamic quantities characterizing an LRS-II spacetime with a perfect fluid at the background level are listed below along with their physical interpretations.
\begin{itemize}
    \item \textbf{Geometric quantities:}
    \begin{itemize}
    \item $\theta \equiv D_{\mu}u^{\mu}$; the expansion of the congruence with the timelike vector $u^{\mu}$.
    \item $\phi \equiv d_{\mu}e^{\mu}$; the expansion of the 2-sheet orthogonal to $u^{\mu},\,e^{\mu}$ along the preferred spatial direction specified by the local rotational symmetry. 
    \item $\Sigma \equiv e^{\mu}e^{\nu}\sigma_{\mu\nu}
= e^{\mu}e^{\nu}D_{\langle\mu}u_{\nu\rangle}
= e^{\mu}e^{\nu} \left(h^{\alpha}_{\,\,\,(\mu}h^{\beta}_{\,\,\,\nu)} - \frac{1}{3}h_{\mu\nu}h^{\alpha\beta}\right)D_{\alpha}u_{\beta}$; 
projection of the shear tensor $\sigma_{\mu\nu}=e^{\mu}e^{\nu}D_{\langle\mu}u_{\nu\rangle}$ along the spacelike direction $e^{\mu}$ \footnote{$\langle...\rangle$ in the indices denotes symmetrization of indices.}.
    \item $\mathcal{E} \equiv E_{\mu\nu}e^{\mu}e^{\nu} 
= C_{\mu\alpha\nu\beta}e^{\mu}e^{\nu}u^{\alpha}u^{\beta}$; projection of the electric part $E_{\mu\nu}$ of the Weyl curvature tensor $C_{\alpha\beta\gamma\delta}$ along the spacelike direction $e^{\mu}$.
    \item $^{3}R$; the 3-curvature of the 3-space orthogonal to $u^{\mu}$.
    \end{itemize}
    \item \textbf{Thermodynamic quantities:}
    \item $\mu \equiv T_{\mu\nu}u^{\mu}u^{\nu}$; energy density of the perfect fluid as measured locally by an observer with the 4-velocity $u^{\mu}$, $T_{\mu\nu}$ being the energy-momentum tensor.
    \item $P \equiv \frac{1}{3}T_{\mu\nu}h^{\mu\nu}$; pressure of the perfect fluid.
\end{itemize}
For elaborate discussions on how these quantities arise in a 1+1+2-decomposition set-up, the reader is referred to the references \cite{Clarkson:2002jz,Clarkson:2007yp}. When the acceleration vector $a^{\mu} = u^{\nu}\nabla_{\nu}u^{\mu}$ for the timelike congruence vanishes, as in the present case under consideration, the 4-vector $u^{\mu}$ satisfies the geodesic equation. Since the time coordinate $t$ in the metric \eqref{LRS-II} is chosen as the affine parameter along $u^{\mu}$, we can identify the timelike congruence with the congruence of comoving geodesics. \footnote{The explicit form of $u^{\mu},\,e^{\mu}$ for the metric \eqref{LRS-II} are $u^{\mu} = (A(t,r),0,0,0),\,\,e^{\mu} = \left(0,\frac{1}{B(t,r)},0,0\right)$.}

The evolution and propagation equations for LRS-II spacetimes with a perfect barotropic fluid ($P=P(\mu)$) are as follows
\begin{itemize}
    \item \textbf{Evolution equations}:
    \begin{subequations}\label{evoln}
    \begin{eqnarray}
    && \dot{\phi} = -\frac{1}{2}\phi\left(\frac{2}{3}\theta - \Sigma\right), \\
    && \dot{\theta} = -\frac{1}{3}\theta^2 - \frac{3}{2}\Sigma^2 - \frac{1}{2}(\mu + 3P(\mu)), \\
    && \dot{\mu} = -\theta(\mu + P(\mu)), \\
    && \dot{\Sigma} = -\frac{1}{2}\Sigma^2 - \frac{2}{3}\theta\Sigma - \mathcal{E}, \\
    && \dot{\mathcal{E}} = - \frac{3}{2}\mathcal{E}\left(\frac{2}{3}\theta - \Sigma\right) - \frac{1}{2}(\mu + P(\mu))\Sigma.
    \end{eqnarray}
    \end{subequations}
    \item \textbf{Propagation equations}:
    \begin{subequations}\label{prop}
    \begin{eqnarray}
    && \hat{\phi} = -\frac{1}{2}\phi^2 + \left(\frac{1}{3}\theta + \Sigma\right)\left(\frac{2}{3}\theta - \Sigma\right) - \frac{2}{3}\mu - \mathcal{E}, \\
    && \hat{\Sigma} - \frac{2}{3}\hat{\theta} = - \frac{3}{2}\phi\Sigma, \\
    && \hat{\mathcal{E}} - \frac{1}{3}\hat{\mu} = - \frac{3}{2}\phi\mathcal{E}, \\
    && \hat{P} = P,_{\mu}\hat{\mu} = 0.
    \end{eqnarray}
    \end{subequations}
    \item \textbf{Friedmann constraint}:
    \begin{equation}\label{fried}
        \frac{\theta^2}{9} + \frac{^{3}R}{6} = \frac{\mu}{3} + \frac{\Sigma^2}{4}.
    \end{equation}
\end{itemize}
The Gaussian curvature $K$ of the 2D sheet is given by $^2R_{\mu\nu} = K N_{\mu\nu}$. The following relation holds \cite{Betschart:2004uu,Goswami:2011ft} between the Gaussian curvature and the 3-curvature
\begin{equation}
    ^3R = -2\left(\frac{1}{2}\hat{\phi} + \frac{3}{4}\phi^2 - K\right)\,.
\end{equation}
Using the $\hat{\phi}$-equation in \eqref{prop}, one can get the following relation
\begin{equation}\label{gauss_curv}
    K = \frac{\mu}{3} - \mathcal{E} + \frac{1}{4}\phi^2 - \left(\frac{\theta}{3} - \frac{\Sigma}{2}\right)^2 \,.
\end{equation}
Its evolution and propagation equations are
\begin{subequations}\label{gauss_curv_eqs}
    \begin{eqnarray}
        && \dot K = -\frac{2}{3}\left(\frac{2}{3}\theta - \Sigma\right)K\,,\\
        && \hat K = - \phi K\,.
    \end{eqnarray}
\end{subequations}
The covariant time and radial derivatives usually do not commute. For a scalar quantity $\psi$, the following commutation relation holds \cite{Singh:2016qmr}
\begin{equation}\label{comm}
        \hat{\dot{\psi}} - \dot{\hat{\psi}} = \left(\frac{1}{3}\theta + \Sigma\right)\hat{\psi}\,.
\end{equation}
An important observation is that, in the absence of the fluid momentum density or anisotropic pressure, the last of the propagation equations implies either a homogeneous matter distribution at all times ($\hat{\mu}=0$) or a fluid with constant pressure. Henceforth we will consider a dust fluid for which the pressure identically vanishes and therefore the last propagation equation is identically satisfied.    
    
The propagation equations \eqref{prop} are spacelike constraints that must hold at all the constant time hypersurfaces. In other words, these constraints must be conserved in time. To show this explicitly, let us define the following combinations:
\begin{subequations}
\begin{eqnarray}
&& \mathcal{C}_1 \equiv \hat{\phi} + \frac{1}{2}\phi^2 - \left(\frac{1}{3}\theta + \Sigma\right)\left(\frac{2}{3}\theta - \Sigma\right) + \frac{2}{3}\mu + \mathcal{E}, \\
&& \mathcal{C}_2 \equiv \hat{\Sigma} - \frac{2}{3}\hat{\theta} + \frac{3}{2}\phi\Sigma, \\
&& \mathcal{C}_3 \equiv \hat{\mathcal{E}} - \frac{1}{3}\hat{\mu} + \frac{3}{2}\phi\mathcal{E}\;.
\end{eqnarray}
\end{subequations}
The propagation equations can be represented as 
\begin{equation}\label{propconst}
    \{\mathcal{C}_1,\mathcal{C}_2,\mathcal{C}_3\} = \{0,0,0\}.
\end{equation}
Using the commutation relation \eqref{comm} we can derive the following equations
\begin{subequations}\label{propevoln}
\begin{eqnarray}
&& \dot{\mathcal{C}}_1 = - \left(\frac{2}{3}\theta + \frac{1}{2}\Sigma\right)\mathcal{C}_1 + \frac{1}{2}\phi\mathcal{C}_2, \\
&& \dot{\mathcal{C}}_2 \equiv - \theta\mathcal{C}_2 - \mathcal{C}_3, \\
&& \dot{\mathcal{C}}_3 = - \frac{1}{2}(\mu + P - 3\mathcal{E})\mathcal{C}_2 - \left(\frac{4}{3}\theta - \frac{1}{2}\Sigma\right)\mathcal{C}_3\;. \\
\end{eqnarray}
\end{subequations}
Eqs. \eqref{propevoln} essentially dictate the evolution of the spacelike constraints given by Eq.\eqref{propconst}. We see that the constraint \eqref{propconst} is conserved in time, i.e. if it is satisfied initially, it is satisfied at all times.

The spacelike constraints in Eq. \eqref{prop} are not purely algebraic as they involve radial derivatives. However, using the commutation relation in Eq. \eqref{comm} it is possible to recast the governing equations as an autonomous system of first-order differential equations with the constraints now being interpreted as purely algebraic ones. The idea is to promote the hat derivatives to separate dynamical variables and calculate their time evolution using the commutation relation Eq. \eqref{comm}. The propagation equations in Eq. \eqref{prop} can then be interpreted as purely algebraic constraints. Hence we get an extended dynamical system (considering dust fluid) given by the following dynamical equations.
\begin{itemize}
    \item 
    \begin{subequations}\label{dynsys_ext}
    \begin{eqnarray}
    && \dot{\phi} = -\frac{1}{2}\phi\left(\frac{2}{3}\theta - \Sigma\right), \\
    && \dot{\theta} = -\frac{1}{3}\theta^2 - \frac{3}{2}\Sigma^2 - \frac{1}{2}\mu, \\
    && \dot{\mu} = -\theta\mu, \\
    && \dot{\Sigma} = -\frac{1}{2}\Sigma^2 - \frac{2}{3}\theta\Sigma - \mathcal{E}, \\
    && \dot{\mathcal{E}} = - \frac{3}{2}\mathcal{E}\left(\frac{2}{3}\theta - \Sigma\right) - \frac{1}{2}\mu\Sigma, \\
    && \dot{\hat{\phi}} = - \left(\frac{2}{3}\theta + \frac{1}{2}\Sigma\right)\hat{\phi} - \frac{1}{2}\phi\left(\frac{2}{3}\hat{\theta} - \hat{\Sigma}\right), \\
    && \dot{\hat{\theta}} = - \theta\hat{\theta} - 3\Sigma\hat{\Sigma} - \frac{1}{2}\hat{\mu} - \Sigma\hat{\theta}, \\
    && \dot{\hat{\mu}} = - \left(\frac{4}{3}\theta + \Sigma\right)\hat{\mu} - \mu\hat{\theta}, \\
    && \dot{\hat{\Sigma}} = - (\theta + 2\Sigma)\hat{\Sigma} - \frac{2}{3}\Sigma\hat{\theta} - \hat{\mathcal{E}}, \\
    && \dot{\hat{\mathcal{E}}} = - \left(\frac{4}{3}\theta - \frac{1}{2}\Sigma\right)\hat{\mathcal{E}} - \frac{3}{2}\mathcal{E}\left(\frac{2}{3}\hat{\theta} - \hat{\Sigma}\right) - \frac{1}{2}\hat{\mu}\Sigma - \frac{1}{2}\mu\hat{\Sigma}\;,
    \end{eqnarray}
    \end{subequations}
\end{itemize}
along with the set of algebraic constraints given by \eqref{prop}. There are three such algebraic constraints in the case of dust matter, allowing us to eliminate $\hat{\phi},\,\hat{\Sigma},\,\hat{\mathcal{E}}$. The reduced dynamical system is: 
\begin{subequations}\label{dynsys}
\begin{eqnarray}
&& \dot{\phi} = -\frac{1}{2}\phi\left(\frac{2}{3}\theta - \Sigma\right), \\
&& \dot{\theta} = -\frac{1}{3}\theta^2 - \frac{3}{2}\Sigma^2 - \frac{1}{2}\mu, \\
&& \dot{\mu} = -\theta\mu, \\
&& \dot{\Sigma} = -\frac{1}{2}\Sigma^2 - \frac{2}{3}\theta\Sigma - \mathcal{E}, \\
&& \dot{\mathcal{E}} = - \frac{3}{2}\mathcal{E}\left(\frac{2}{3}\theta - \Sigma\right) - \frac{1}{2}\mu\Sigma, \\
&& \dot{\hat{\theta}} = - \theta\hat{\theta} - 3\Sigma\hat{\theta} + \frac{9}{2}\Sigma^2\phi - \frac{1}{2}\hat{\mu}, \\
&& \dot{\hat{\mu}} = -\left(\frac{4}{3}\theta + \Sigma\right)\hat{\mu} - \mu\hat{\theta}.
\end{eqnarray}
\end{subequations}

\subsection{Some discussion about the structure of the system}\label{subsec: system_structure}

It is worthwhile here to ponder upon the structure of the system of equations \eqref{dynsys}. The system \eqref{dynsys} by itself is a closed set, but one can notice a nice decoupling here. The $\{\dot\theta,\dot\mu,\dot\Sigma,\dot{\mathcal{E}}\}$-equations are decoupled from the other equations. $\theta,\,\mu,\,\Sigma,\,\mathcal{E}$ by themselves form a closed set of equations since they do not contain the other covariant quantities. This system is well-posed \cite{Coley:2008qd}. Once one knows $\theta,\,\mu,\,\Sigma,\,\mathcal{E}$, either in the form of fixed points or as numerical solutions, one can plug them into the $\{\dot\phi,\dot{\hat\theta},\dot{\hat{\mu}}\}$-equations and solve for $\phi(t),\,\hat\theta(t),\,\hat\mu(t)$. The propagation equations \eqref{prop} then allow us to find out $\hat{\phi}$, $\hat{\Sigma}$ and $\hat{\mathcal{E}}$.

We are considering here a congruence of comoving geodesics. Let us follow the geodesic of a particular comoving observer and consider a covariant quantity $Q$. If we go to the tangent space of a comoving observer, because of the inhomogeneity of the system, the observer has at his disposal two derivatives; $\dot{Q}$ and $\hat{Q}$. The $\dot{Q}$ gives the change in the value of $Q$ at his tangent space, i.e. along his particular geodesic, while $\hat{Q}$ can be interpreted as a ``radial perturbation'' of $Q$ on the same constant time hypersurface, where time is measured along the unique timelike 4-vector field $u^{\mu}$ \footnote{This is not a generic spatial perturbation, though, but a specific kind of perturbations respecting the local rotational symmetry.}. This can be seen as follows. Consider a comoving observer at $r=r_0$, who, at a particular time $t_0$, has the value $Q=Q_0$. Now consider a neighboring comoving observer at $r = r_0 + \delta r$. Since both the observers are comoving observers, their geodesics belong to the same congruence. The deviation vector between their geodesics is given by $\xi^{\mu} = \delta r e^{\mu}$. The value of at the time $t_0$ for this neighbouring observer is given by
\begin{equation}
    Q = Q_0 + \delta_{\xi}Q = Q_0 + \xi^{\mu}\nabla_{\mu}Q = Q_0 + \delta r e^{\mu}(- \dot{Q}u_{\mu} + \hat{Q}e_{\mu} + N^{\,\,\,\nu}_{\mu}D_{\nu}Q) = Q_0 + \hat{Q}\delta r\,,
\end{equation}
where in the last step we utilized the orthogonality of $e^{\mu}$ with $u^{\mu}$ and $N_{\mu\nu}$. With respect to the comoving observer at $r=r_0$, the non-vanishing $\delta_{\xi}Q=\hat{Q}\delta r$ is precisely what is stopping the other comoving observers from having the same $Q$-value $Q_0$, making the spatial distribution of $Q$ inhomogeneous. This justifies the interpretation of $\hat{Q}$ as a perturbation.
The $\dot{\hat{Q}}$-equation gives the dynamics of this radial perturbation. 

$\theta,\,\Sigma,\,\mathcal{E},\,\mu$ are the expansion, shear, electric part of the Weyl tensor and the energy density locally measured by a comoving observer. On the other hand, the covariant quantity $\phi$ is related to the Gaussian curvature of a shell (Eq. \eqref{gauss_curv}), which characterizes the ``shape'' of the shells. The sign of the Gaussian curvature distinguishes between spherically, cylindrically and hyperbolically symmetric spatial geometry. For spherically and cylindrically symmetric situations, different comoving shells will in general have different values of the Gaussian curvature. When this is the case, $\phi$ can be thought of as labels for different shells. 

The decoupling of the $\{\dot\theta,\,\dot\Sigma,\,\dot{\mathcal{E}},\,\dot\mu\}$-equations from the other equations imply that the qualitative nature of the evolution of expansion, shear, electric part of the Weyl tensor and the energy density is the same for all the comoving observers and is independent of how the radial perturbations in the respective quantities evolve. However, the existence of such radial perturbations implies that the evolution of different shells will fall ``out of phase''. Different shells will not experience the same evolutionary phase simultaneously. This is the key contrast with a homogeneous geometry e.g. FLRW, where \emph{all} the shells will experience the same evolutionary phase simultaneously.
 
In general, when the fluid is not perfect, there will be mixed equations and the radial perturbations do affect the shell dynamics. In such cases, $\{\dot\theta,\dot\mu,\dot\Sigma,\dot{\mathcal{E}}\}$-equations cannot be decoupled from the other equations.

It should be clear from the above discussion that if $Q$ is a covariant quantity, then in our approach $Q=0$ does \emph{not} automatically imply $\hat{Q}=0$. Each comoving observer has a 4D tangent space. Consider locally the 2D projection of the tangent space of a comoving observer spanned by the 4-vectors $u^{\mu}$ and $e^{\mu}$ (the local $u$-$e$ plane). In the local $u$-$e$ plane of the comoving observer, $Q=0$ corresponds to a curve. The normal vector to this curve is given by 
\begin{equation}
    \nabla_{\mu}Q = - \dot{Q}u_{\mu} + \hat{Q}e_{\mu} + N^{\nu}_{\mu}D_{\nu}Q\,.
\end{equation}
The last term, $N^{\nu}_{\,\,\,\mu}D_{\nu}Q$ is the gradient of $Q$ on the 2-sheet orthogonal to $u^{\mu}$ and $e^{\mu}$. Since in the 1+1+2 decomposition of LRS spacetimes the spacelike vector field $e^{\mu}$ at each spatial point is taken to be along the axis of local rotational symmetry, the last term identically vanishes. Let the tangent to the curve $Q=0$ in the local $u$-$e$ plane be given by 
\begin{equation}
    Q^{\mu} = \alpha u^{\mu} + \beta e^{\mu}.
\end{equation}
The condition $Q^{\mu}\nabla_{\mu}Q=0$ helps to determine the slope to this curve in the local $u$-$e$ plane. 
\begin{equation}
    - \frac{\alpha}{\beta} = \frac{\hat{Q}}{\dot{Q}}.
\end{equation}
If this slope is positive, then $Q=0$ is attained by a comoving observer at a larger radius at a later time and by a comoving observer at a smaller radius at a smaller time. The opposite is true when the slope is negative. In special cases when $\hat{Q}$ also vanishes along the $Q=0$ curve, the slope is zero and all comoving observers attain $Q=0$ at the same time. In this case, the $Q$ distribution is homogeneous.\footnote{Which does not mean that the spacetime geometry as a whole should be homogeneous. The hat of other covariant quantities can be non-vanishing.}


\section{A dynamical system formulation for LRS-II spacetime with dust}\label{sec: lrs2_ps}

Throughout the rest of the paper, we specialize to that case of pressureless dust. The first step towards a dynamical system analysis is to define dimensionless dynamical variables:
\begin{equation}\label{dynvar_def}
    \begin{aligned}
    & x_1\equiv\frac{\phi}{\theta}, \quad x_2\equiv\frac{3\mu}{\theta^2}, \quad x_3\equiv\frac{3}{2}\frac{\Sigma}{\theta}, \quad x_4\equiv\frac{\mathcal{E}}{\theta^2},\\
    & y_1\equiv\frac{\hat{\phi}}{\theta^2}, \quad y_2\equiv\frac{3\hat{\mu}}{\theta^3}, \quad y_3\equiv\frac{3}{2}\frac{\hat{\Sigma}}{\theta^2}, \quad y_4\equiv\frac{\hat{\mathcal{E}}}{\theta^3},\\
    & z\equiv\frac{\hat{\theta}}{\theta^2}.
    \end{aligned}
\end{equation}
In terms of these variables the Friedmann constraint \eqref{fried} is
\begin{equation}\label{fried_dimless}
        x_2 + x_3^2 = 1 + \frac{3}{2}\frac{^{3}R}{\theta^2}.
    \end{equation}
Next we introduce a dimensionless time variable $\tau$ as $d\tau=\epsilon\theta dt$, where $\epsilon=+1$ for an expanding LRS-II spacetime ($\theta>0$) and $\epsilon=-1$ for a contracting LRS-II spacetime ($\theta<0$). We imported the $\epsilon$ so that $\tau$ is always a monotonically increasing function of time, and therefore can be justifiably taken as a time variable on the phase space. So the dynamical system formulation we develop here can be applied to study both inhomogeneous dark energy models and the phenomenon of gravitational collapse.

An important relationship that is worth writing here is
\begin{equation}\label{imp_rel}
    \frac{\dot{\theta}}{\theta^2} = - \frac{1}{6}(2 + x_2 + 4x_3^2).
\end{equation}
Considering a dust fluid ($P(\mu)=0$) and using the equations \eqref{dynsys_ext}, we get the following dynamical system
\begin{subequations}\label{dynsys_ext_dimless}
\begin{eqnarray}
&& \frac{dx_1}{\epsilon d\tau} = \frac{1}{6}x_1 (2x_3 + x_2 + 4x_3^2), \\
&& \frac{dx_2}{\epsilon d\tau} = - \frac{1}{3}x_2 (1 - x_2 - 4x_3^2), \\
&& \frac{dx_3}{\epsilon d\tau} = - \frac{1}{6}x_3 (2 + 2x_3 - x_2 - 4x_3^2) - \frac{3}{2}x_4, \\
&& \frac{dx_4}{\epsilon d\tau} = - \frac{1}{3}x_4 (1 - 3x_3 - x_2 - 4x_3^2) - \frac{1}{9}x_2 x_3, \\
&& \frac{dy_1}{\epsilon d\tau} = \frac{1}{3}y_1 (x_2 - x_3 + 4x_3^2) + \frac{1}{3}x_1(y_3 - z), \\
&& \frac{dy_2}{\epsilon d\tau} = - \frac{1}{3}y_2 + \frac{1}{6}y_2(3x_2 - 4x_3) - x_{2}z + 2x_3^{2}y_2, \\
&& \frac{dy_3}{\epsilon d\tau} = - \frac{1}{3}y_3 - \frac{3}{2}y_4 + \frac{1}{3}y_3(x_2 - 4x_3) - \frac{2}{3}x_{3}z + \frac{4}{3}x_3^{2}y_3, \\
&& \frac{dy_4}{\epsilon d\tau} = - \frac{1}{3}y_4 - \frac{1}{18}x_2(2y_3 - 9y_4) - \frac{1}{9}x_3(y_2 - 3y_4) + x_4(y_3 - z) + 2x_3^{2}y_4, \\
&& \frac{dz}{\epsilon d\tau} = - \frac{1}{6}y_2 - \frac{1}{3}z + \frac{1}{3}x_{2}z - \frac{2}{3}x_3(2y_3 + z) + \frac{4}{3}x_3^{2}z.
\end{eqnarray}
\end{subequations}
The propagation equations \eqref{prop} can now be written explicitly as algebraic constraints
\begin{subequations}\label{prop_dimless}
\begin{eqnarray}
&& y_1 = - \frac{1}{2}x_1^2 + \frac{2}{9}(1-x_3)(1+2x_3) - \frac{2}{9}x_2 - x_4, \\
&& y_3 - z = - \frac{3}{2} x_{1}x_{3}, \\
&& y_4 - \frac{1}{9}y_2 = - \frac{3}{2}x_{1}x_{4}
\end{eqnarray}
\end{subequations}
Using the three algebraic constraints in Eq.\eqref{prop_dimless} to eliminate $y_1,\,y_3,\,y_4$ one gets a reduced dynamical system which corresponds to the reduced system of equations \eqref{dynsys}
\begin{subequations}\label{dynsys_dimless}
\begin{eqnarray}
&& \frac{dx_1}{\epsilon d\tau} = \frac{1}{6}x_1 (2x_3 + x_2 + 4x_3^2), \\
&& \frac{dx_2}{\epsilon d\tau} = - \frac{1}{3}x_2 (1 - x_2 - 4x_3^2), \\
&& \frac{dx_3}{\epsilon d\tau} = - \frac{1}{6}x_3 (2 + 2x_3 - x_2 - 4x_3^2) - \frac{3}{2}x_4, \\
&& \frac{dx_4}{\epsilon d\tau} = - \frac{1}{3}x_4 (1 - 3x_3 - x_2 - 4x_3^2) - \frac{1}{9}x_2 x_3, \\
&& \frac{dy_2}{\epsilon d\tau} = - \frac{1}{3}y_2 + \frac{1}{6}y_2(3x_2 - 4x_3) - x_{2}z + 2x_3^{2}y_2, \\
&& \frac{dz}{\epsilon d\tau} = - \frac{1}{6}y_2 - \frac{1}{3}z + \frac{1}{3}x_{2}z - 2x_{3}z - 2x_{1}x_3^2 + \frac{4}{3}x_3^{2}z.
\end{eqnarray}
\end{subequations}

\subsection{Interpretation of the dynamical structures}\label{subsec: interpretation}

The approach that we are going to take here is the following. Since, in a generically inhomogeneous situation, the dynamics with respect to observers at different comoving shells are not equivalent, we will go to the frame of a particular comoving observer and interpret the dynamical structures, namely fixed points and invariant submanifolds, from that observer's point of view. As explained before, a particular comoving observer has, at his disposal the covariant geometrical and thermodynamic quantities as well as their radial gradients as given by the hat derivatives. Both these sets have to be taken into account to get a full picture of the dynamics. The dynamics are described in the 6D phase space $\{x_1,x_2,x_3,x_4,y_2,z\}$ by the phase flow given by the autonomous system \eqref{dynsys_dimless}.

In general, if the comoving observer at $r=r_0$ is undergoing an evolutionary phase characterized by a particular fixed point, a neighbouring observer at $r=r_0+\delta r$ need not necessarily undergo the same evolutionary phase, because of generally non-vanishing radial perturbations. This can be seen more clearly as follows. If $\{x_1^*,x_2^*,x_3^*,x_4^*,y_2^*,z^*\}$ are the coordinates of a fixed point, then the fixed point is characterized by the vanishing of the following covariant quantities: 
\begin{equation*}
    \{\phi-x_1^*\theta,3\mu-x_2^*\theta^2,3\Sigma-2x_3^*\theta,\mathcal{E}-x_4^*\theta^2,3\hat{\mu}-y_2^*\theta^3,\hat{\theta}-z^*\theta^2\} = \{0,0,0,0,0,0\}\,.
\end{equation*}
According to the discussion at the end of section \ref{subsec: system_structure}, the vanishing of a covariant quantity $Q$ does not automatically imply the vanishing of its radial perturbation $\hat{Q}$. This gives rise to a non-vanishing slope of the $Q=0$ curve in the local $u$-$e$ plane of a comoving observer, which means that a neighbouring comoving observer at $r=r_0+\delta r$ will attain the same fixed point at a slightly earlier or a slightly later time. The dynamics of the radial perturbations with respect to a comoving observer, in a covariant manner, are given by the $y$-equations.

For a certain fixed point, if $\{y_1,y_2,y_3,y_4,z\}=\{0,0,0,0,0\}$, then that implies $\{\hat{\phi},\hat{\theta},\hat{\Sigma},\hat{\mathcal{E}},\hat{\mu}\}=\{0,0,0,0,0\}$; the radial perturbations in all the covariant quantities of interest vanishes. In such a scenario, observers at different comoving shells measure the same matter density and experience the same spacetime geometry. Such a fixed point necessarily corresponds to a completely homogeneous scenario. An example of such a geometry is the Kantowski-Sachs spacetime. However, the converse is not necessarily true. For a fixed point to represent a homogeneous geometry and a homogeneous matter distribution, not all the hat derivatives should necessarily vanish. The prime example for such a case is the FLRW geometry, for which $\hat\phi\neq0$ but the hat derivatives of all the other covariant geometric and thermodynamic quantities vanish.

The above discussion regarding the fixed points also extends to points on invariant submanifolds, as they are also characterized by the vanishing of certain covariant quantities.


\section{Invariant submanifolds}\label{sec: inv_sub}

In this section, we present the invariant submanifolds. Each of them divides the entire phase space into two disjoint sectors. The interpretations of these invariant submanifolds are given assuming that the expansion $\theta$ remains finite. For a singular situation when $\theta\rightarrow\pm\infty$, the below interpretations need not necessarily hold true.

The invariant submanifolds containing inhomogeneous solutions are as follows:

\begin{enumerate}

    \item $\mathcal{S}_\phi$: The invariant submanifold $x_1=0$ corresponds to $\phi=0$. Recalling that $\phi$ measures the expansion of the 2D sheet along the spacelike direction $e^{\mu}$, this invariant submanifold contains all the solutions where a comoving observer measures zero expansion of the 2D sheet along that direction. For the metric \eqref{LRS-II}, it can be calculated that
    \begin{equation}\label{phi_lrs2}
        \phi = \frac{1}{B(t,r)}\left(2\frac{C_{,r}(t,r)}{C(t,r)}-\frac{A_{,r}(t,r)}{A(t,r)}\right)
    \end{equation}
    This might include solutions at spatial infinity $B(t,r)\rightarrow\infty$ (since at spatial infinity one can consider the radial lines to be parallel to each other) and also Kantowski-Sachs solutions $A=1,\,B=B(t),\,C=C(t),\,D(\theta)=\sin\theta$. 
    
    The stability of this invariant submanifold depends on the sign of the quantity  
    \begin{equation}
        2x_3 + x_2 + 4x_3^2 = \frac{3\Sigma}{\theta}\left(1 + \frac{3\Sigma}{\theta}\right) + \frac{3\mu}{\theta^2}.
    \end{equation}
    
    \item $\mathcal{S}_K$: This is the submanifold of vanishing Gaussian curvature of the 2D sheet. Let us define \begin{equation}\label{gauss_curv_dimless}
        x_K \equiv \frac{9K}{\theta^2} = x_2 - 9x_4 + \frac{9}{4}x_1^2 - (1 - x_3)^2\,,
    \end{equation}
    where for the second equality we have utilized the expression \eqref{gauss_curv}. It is easy to find out the evolution equation for $x_K$ using the evolution equation for Gaussian curvature in \eqref{gauss_curv_eqs}: 
    \begin{equation}
        \frac{dx_K}{\epsilon d\tau} = \frac{1}{9}x_K (2 + 3x_2 + 4x_3 + 12x_3^2)\,.
    \end{equation}
    For the metric \eqref{LRS-II}, it can be calculated that
    \begin{equation}\label{K_lrs2}
        K = \frac{1}{C^{2}(t,r)}\frac{D''(y)}{D(y)} = 
        \begin{cases}
    -\frac{1}{C^{2}(t,r)}, \qquad k=+1 \\
    0, \qquad k=0 \\
    \frac{1}{C^{2}(t,r)}, \qquad k=-1
    \end{cases}\,.
    \end{equation}
    
    This invariant submanifold, which contains cylindrically symmetric solutions $D(y)=y$, divides the entire phase space into two disjoint subsections, containing solutions of positive Gaussian curvature (i.e. spherically symmetric solutions) and negative Gaussian curvature (i.e. hyperbolically symmetric solutions). Apart from cylindrically symmetric solutions, it is also possible for $\mathcal{S}_K$ to contain spherically or hyperbolically symmetric solutions for a comoving observer at an infinite spatial distance away from the origin, which is taken to be the symmetry centre\footnote{For homogeneous spacetimes, there is no unique symmetry centres. The origin can be placed anywhere.}. This is because, even for spherically and hyperbolically symmetric solutions, the Gaussian curvature of the shells will decrease as they are further away from the origin. Existence of $\mathcal{S}_K$ means that the Gaussian curvature of the 2D sheets cannot change sign. 
    
    The stability of this invariant submanifold depends on the quantity 
    \begin{equation}
        2 + 3x_2 + 4x_3 + 12x_3^2 = 2 + 6\frac{\Sigma}{\theta} + \frac{9\mu}{\theta^2} + \frac{27\Sigma^2}{\theta^2}\,.
    \end{equation}
        
    \item $\mathcal{S}_\mu$: The invariant submanifold $x_2=0$ corresponds to the vacuum scenario\footnote{In presence of matter there was a unique choice for the timelike unit 4-vector $u^{\mu}$ that was the basis of the 1+3-decompositiion, namely, the unit vector tangent to the worldline of the matter particles. In vacuum, we do not have that choice. For cosmology, we can choose the ``frame'', i.e. the unit 4-vector $u^{\mu}$ such that the CMB dipole moment vanishes.\cite{Ellis:1998ct}. More discussion on this important issue is presented at the beginning of the next section.}. On this submanifold the Friedmann equation \eqref{fried_dimless} becomes
    \begin{equation}
            x_3^2 = 1 + \frac{3}{2}\frac{^{3}R}{\theta^2}.
            \end{equation} 
    As expected, the vacuum case represents an invariant submanifold because there is no matter production mechanism in classical physics. The stability of this invariant submanifold depends on the sign of the quantity
    \begin{equation}
        1 - 4x_3^2 = 1 - \frac{9\Sigma^2}{\theta^2} = -3\left(1 + 2\frac{^{3}R}{\theta^2}\right).
    \end{equation}
    Following are some of the generic conclusions regarding the stability of this invariant submanifold that we can reach from the above expression:
    \begin{itemize}
        \item When shear is negligible ($|\Sigma/\theta|\ll1$), $\mathcal{S}_\mu$ is of attracting nature for and expanding LRS-II ($\epsilon=+1$) and of repelling nature for a contracting LRS-II ($\epsilon=-1$). This is consistent with the intuitive conclusion that the any pre-existing small amount of matter density must dilute away with an expansion whereas it must go up with a contraction. Note that this is in general not true in presence of non-negligible shear.
        \item For spatially flat or positively curved LRS-II ($^{3}R\geq0$), $\mathcal{S}_\mu$ is always of repelling nature for an expanding LRS-II ($\epsilon=+1$) and of attracting nature for a contracting LRS-II ($\epsilon=-1$). One may be tempted to interpret this result directly as any pre-existing small amount of matter density is actually going up with an expansion and diluting with contraction. This sounds counter-intuitive and one can clearly see from the $\dot{\mu}$-equation in the system \eqref{dynsys}, matter density definitely does go down with an expansion and go up with a contraction. Indeed, it would be a wrong interpretation. Instead, the correct interpretation should be that any pre-existing small amount of matter density becomes dominant over the shear with expansion and becomes subdominant with respect to the shear with contraction. This behaviour is similar to what one expects for the case of homogeneous and anisotropic Bianchi spacetime in presence of an isotropic fluid, where the matter density goes as $\sim\frac{1}{a^3}$ and metric anisotropy goes as $\sim\frac{1}{a^6}$. This same behaviour is preserved even when inhomogeneity is introduced in both the metric and the matter sectors. 
        
        Note that one cannot conclude the same for spatially negatively curved LRS-II.
        \item The above two conclusions are not in conflict as the isotropic vacuum limit can only be supported with negative spatial curvature. The isotropic vacuum limit with negative spatial curvature can be interpreted as an inhomogeneous version of the well-known Milne solutions.
    \end{itemize}
   
   As explained at the end of the last section, a particular comoving observer being on the invariant submanifold $\mathcal{S}_{\mu}$ does not mean that a neighbouring comoving observer will also be on the same invariant submanifold. In other words, even if a particular comoving observer experiences vacuum locally, a neighbouring observer can experience locally a non-vanishing energy density. This is because of a generally non-vanishing $\hat{\mu}$, which is interpreted as some radial perturbations in $\mu$. The evolution equation of this radial perturbation can be obtained from the last two equations of \eqref{dynsys}
    \begin{equation}\label{ptbn_1}
        \Ddot{\hat{\mu}} + \left(\frac{4}{3}\theta + \Sigma\right)\dot{\hat{\mu}} - \left(\frac{4}{9}\theta^2 + \frac{5}{2}\Sigma^2 + \frac{2}{3}\theta\Sigma + \mathcal{E}\right)\hat{\mu} = 0\,.
    \end{equation}
    However, it is more illuminating to write the evolution equation for the dimensionless quantity $y_2$ here, 
    \begin{equation}
        \frac{dy_2}{\epsilon d\tau} = - \frac{1}{3} y_2 (1 + 2x_3 - 6x_3^2)\,.
    \end{equation}
    Whether the radial perturbations decay ($y_2 \rightarrow0$) or not during an expanding or a contracting LRS-II, depends on the sign of the quantity $(1 + 2x_3 - 6x_3^2)$. If the radial perturbation decays, the situation asymptotically tends to be vacuum for \emph{all} the observers. 

    \item $\mathcal{S}_{\Sigma\mathcal{E}}$: The invariant submanifold $\{x_3,x_4\}=\{0,0\}$ corresponds to $\{\Sigma,\mathcal{E}\}=\{0,0\}$. On this submanifold the Friedmann equation \eqref{fried_dimless} becomes
    \begin{equation}
            x_2 = 1 + \frac{3}{2}\frac{^{3}R}{\theta^2}.
            \end{equation}
    The stability of this invariant submanifold can be investigated from the Jacobian
    \begin{equation*}
        \left(\begin{smallmatrix}
        \frac{\partial}{\partial x_3}\left(\frac{dx_3}{d\tau}\right) & \frac{\partial}{\partial x_4}\left(\frac{dx_3}{d\tau}\right) \\ 
        \frac{\partial}{\partial x_3}\left(\frac{dx_4}{d\tau}\right) & \frac{\partial}{\partial x_4}\left(\frac{dx_4}{d\tau}\right)
        \end{smallmatrix}\right)_{x_3=0,x_4=0},
    \end{equation*}
    whose eigenvalues are $\frac{\epsilon}{12}\left(3x_2 - 4 \pm \sqrt{x_2^2 + 24x_2}\right)$. Following are some of the generic conclusions that can be drawn:
        \begin{itemize}
            \item In a vacuum ($x_2=0$), $\mathcal{S}_{\Sigma\mathcal{E}}$ is always of attracting (repelling) nature for an expanding (contracting) LRS-II.
            \item In the presence of matter, the stability of $\mathcal{S}_{\Sigma\mathcal{E}}$ depends entirely on the matter density $\mu$ (or, equivalently, the 3-curvature $^{3}R$).   
            \begin{itemize}
                \item For an expanding LRS-II spacetime, $\mathcal{S}_{\Sigma\mathcal{E}}$ is attracting when $0 \leq x_2 < 3-\sqrt{7}$.
                \item For a contracting LRS-II spacetime, $\mathcal{S}_{\Sigma\mathcal{E}}$ is attracting when $x_2 > 3+\sqrt{7}$.
                \item For a spatially flat ($^{3}R=0$) expanding or contracting LRS-II, there will always be one direction in the phase space along which this invariant submanifold is attracting, and one direction along which it is repelling.
            \end{itemize}
        \end{itemize}

    For the special case of a spherically symmetric spacetime (i.e. the Gaussian curvature $K>0$), $\{\Sigma,\mathcal{E}\}=\{0,0\}$ implies a locally FLRW spacetime. A comoving observer passing through an evolutionary phase with $\{\Sigma,\mathcal{E}\}=\{0,0\}$ (i.e. on the invariant submanifold $\{x_3,x_4\}=\{0,0\}$) experiences locally an FLRW spacetime. However, the observer will experience, in general, non-vanishing radial perturbations in the spacetime geometry and matter density, as given by the quantities $\{\hat{\theta},\hat{\Sigma},\hat{\mathcal{E}},\hat{\mu}\}$. The evolution equations for $\{\hat{\theta},\hat{\mu}\}$, can be obtained from the last two equations of \eqref{dynsys}
    \begin{subequations}\label{ptbn_2}
        \begin{eqnarray}
            && \Ddot{\hat{\theta}} + \theta\dot{\hat{\theta}} - \left(\frac{1}{3}\theta^2 + \mu\right)\hat{\theta} = \frac{2}{3}\theta\hat{\mu}\,,\\
            && \Ddot{\hat{\mu}} + \frac{4}{3}\theta\dot{\hat{\mu}} - \left(\frac{4}{9}\theta^2 + \frac{7}{6}\mu\right)\hat{\mu} = 2\mu\theta\hat{\theta}\,.
        \end{eqnarray}
    \end{subequations}
    Subsequently, $\{\hat{\Sigma},\hat{\mathcal{E}}\}$ can be found from Eq.\eqref{prop}. These non-vanishing radial perturbations will stop the geometry from being totally homogeneous. If a comoving observer at $r=r_*$ experiences locally an FLRW geometry, a neighbouring comoving observer $r_1=r_*+\delta r$ will not, in general, experience an FLRW geometry. Similar arguments apply for the cases of cylindrically and hyperbolically symmetric spacetimes.
       
\end{enumerate}

The points on the invariant submanifolds $\mathcal{S}_{\phi},\,\mathcal{S}_K,\,\mathcal{S}_\mu,\,\mathcal{S}_{\Sigma\mathcal{E}}$ correspond, in general, to inhomogeneous solutions. Take, for example, the invariant submanifold $\mathcal{S}_\mu$. If a shell is initially on $\mathcal{S}_\mu$, then it will always stay on it. Physically, if the matter density $\mu$ vanishes on a particular shell, then it will always remain vanishing on that shell forever. At the same time, the other shells will, in general, have a nonzero matter density, and their matter densities will also change with time. Similar interpretation goes for the invariant submanifolds $\mathcal{S}_\phi$ and $\mathcal{S}_{\Sigma\mathcal{E}}$. 

Below we present a special invariant submanifold, on which all the points correspond to homogeneous solutions.

\begin{enumerate}

    \item $\mathcal{S}_{\rm hom}$: A more restricted invariant submanifold is given by $\{x_3,x_4,y_2,z\}=\{0,0,0,0\}$, which corresponds to homogeneous solutions. The points on this invariant submanifold that lie within the region $x_K>0$, which correspond to spherically symmetric solutions, represent the homogeneous FLRW geometry. The points on this submanifold that lie within the region $x_K\leq0$ represent homogeneous geometries that are hyperbolically or cylindrically symmetric counterparts of FLRW geometry. 
    
    As one can see from the dynamical system \eqref{dynsys_dimless}, inhomogeneity creeps into the dynamics through the dynamical variables $y_2$ and $z$. Note that, by itself, $\{y_2,z\}=\{0,0\}$ is \emph{not} an invariant submanifold. Even if we start initially with a homogeneous geometry ($z=0$) and a homogeneous matter distribution ($y_2=0$), non-vanishing sheet expansion and shear ($x_1 x_3\neq0$) can induce inhomogeneity in the geometry ($z$), which, in turn, induces inhomogeneity in the matter distribution ($y_2$) by virtue of the $x_{2}z$ term. However, if we start with the invariant sub-manifold $\{x_3,x_4,y_2,z\}=\{0,0,0,0\}$, then there is no way for inhomogeneity to creep into the system. This implies that a homogeneous universe stays homogeneous in the absence of any radial perturbations, which is consistent with $\mathcal{S}_{\rm hom}$ being an invariant submanifold.

    Note the difference between $\mathcal{S}_{\Sigma\mathcal{E}}$ and $\mathcal{S}_{\rm hom}$. $\mathcal{S}_{\Sigma\mathcal{E}}$ had vanishing shear $\Sigma$ and electric Weyl tensor $\mathcal{E}$ ($\{x_3,x_4\}=\{0,0\}$), but that did not automatically imply vanishing $\hat{\Sigma}$ and $\hat{\mathcal{E}}$ ($\{y_3,y_4\}\neq\{0,0\}$). Not all the comoving shells were experiencing an evolutionary phase with vanishing shear and electric Weyl tensor simultaneously. Therefore, the solutions in $\mathcal{S}_{\Sigma\mathcal{E}}$ were inhomogeneous solutions. On the other hand, $\mathcal{S}_{\rm hom}$ also has vanishing shear $\Sigma$ and electric Weyl tensor $\mathcal{E}$ ($\{x_3,x_4\}=\{0,0\}$), and by virtue of also having $\{y_2,z\}=\{0,0\}$, the constraint equations \eqref{prop_dimless} guarantee vanishing $\{\hat{\Sigma}$ and $\hat{\mathcal{E}}\}$ ($\{y_3,y_4\}\neq\{0,0\}$). In this case, \emph{all} the comoving observers experience an evolutionary phase with vanishing shear and electric Weyl tensor, which is precisely the case of a homogeneous FLRW spacetime geometry. The invariant submanifold $\mathcal{S}_{\rm hom}$ is actually a subset of $\mathcal{S}_{\Sigma\mathcal{E}}$: $\mathcal{S}_{\rm hom}\subset\mathcal{S}_{\Sigma\mathcal{E}}$, that is singled out by the vanishing of the radial perturbations.

     For a spherically symmetric solution on the invariant submanifold $\mathcal{S}_{\Sigma\mathcal{E}}$, the local spacetime geometry with respect to the comoving observer at $r=r_*$ is FLRW. But because of the non-vanishing radial perturbations, a neighbouring comoving observer at $r_1=r_*+\delta r$ will not, in general, experience an FLRW geometry. For the invariant submanifold $\mathcal{S}_{\rm hom}$, there are no such radial perturbations. If the comoving observer at $r_0=r_*$ experiences an FLRW geometry, a neighbouring comoving observer at $r_1=r_*+\delta r$ will also experience the same FLRW geometry. This argument is inductive. One can then focus on this observer at $r_1=r_*+\delta r$ and apply the same argument. By the principle of induction, this argument can be extended to \emph{all} the comoving observers, which implies that all the comoving observers experience the same local FLRW geometry. Similar argument applies to cylindrically and hyperbolically symmetric spacetimes.

The stability of this invariant submanifold can be investigated from the Jacobian
    \begin{equation*}
        \left(\begin{smallmatrix}
        \frac{\partial}{\partial x_3}\left(\frac{dx_3}{d\tau}\right) & \frac{\partial}{\partial x_4}\left(\frac{dx_3}{d\tau}\right) & \frac{\partial}{\partial y_2}\left(\frac{dx_3}{d\tau}\right) & \frac{\partial}{\partial z}\left(\frac{dx_3}{d\tau}\right) \\ 
        \frac{\partial}{\partial x_3}\left(\frac{dx_4}{d\tau}\right) & \frac{\partial}{\partial x_4}\left(\frac{dx_4}{d\tau}\right) & \frac{\partial}{\partial y_2}\left(\frac{dx_4}{d\tau}\right) & \frac{\partial}{\partial z}\left(\frac{dx_4}{d\tau}\right) \\
        \frac{\partial}{\partial x_3}\left(\frac{dy_2}{d\tau}\right) & \frac{\partial}{\partial x_4}\left(\frac{dy_2}{d\tau}\right) & \frac{\partial}{\partial y_2}\left(\frac{dy_2}{d\tau}\right) & \frac{\partial}{\partial z}\left(\frac{dy_2}{d\tau}\right) \\
        \frac{\partial}{\partial x_3}\left(\frac{dz}{d\tau}\right) & \frac{\partial}{\partial x_4}\left(\frac{dz}{d\tau}\right) & \frac{\partial}{\partial y_2}\left(\frac{dz}{d\tau}\right) & \frac{\partial}{\partial z}\left(\frac{dz}{d\tau}\right) \\
        \end{smallmatrix}\right)_{x_3=0,x_4=0,y_2=0,z=0},
    \end{equation*}
whose eigenvalues are
\begin{equation*}
    \frac{\epsilon}{12}\left(3x_2 - 4 \pm \sqrt{x_2^2 + 24x_2}\right),\,\,\frac{\epsilon}{12}\left(5x_2 - 4 \pm \sqrt{x_2^2 + 24x_2}\right)\;.
\end{equation*}
The first pair of eigenvalues are exactly the ones we obtained for $\mathcal{S}_{\Sigma\mathcal{E}}$, whereas it can be checked that the eigenvectors corresponding to the last pair of eigenvalues lie entirely on $\mathcal{S}_{\Sigma\mathcal{E}}$. Since $\mathcal{S}_{\rm hom}\subset\mathcal{S}_{\Sigma\mathcal{E}}$, we conclude that a part of the stability of $\mathcal{S}_{\rm hom}$ comes from the flow in the vicinity of $\mathcal{S}_{\Sigma\mathcal{E}}$, which is given by the first pair of eigenvalues. The other part of the stability of $\mathcal{S}_{\rm hom}$ comes actually from the flow within $\mathcal{S}_{\Sigma\mathcal{E}}$, which is given by the last pair of eigenvalues. This last pair of eigenvalues are essentially related to the dynamics of the radial perturbations on $\mathcal{S}_{\Sigma\mathcal{E}}$. When both of them are negative, one can say that the radial perturbations tend to die out, driving the geometry towards a globally homogeneous one. 

The stability of this invariant submanifold dictates the evolution of some LRS-II geometries that are almost homogeneous. Following are some of the generic conclusions that can be drawn:
\begin{itemize}
     \item In a vacuum ($x_2=0$), $\mathcal{S}_{\rm hom}$ is always of attracting (repelling) nature for an expanding (contracting) LRS-II.
     \item In presence of matter, the stability of $\mathcal{S}_{\rm hom}$ depends entirely on the matter density $\mu$ (or, equivalently, the 3-curvature $^{3}R$).   
        \begin{itemize}
            \item For an expanding LRS-II spacetime, $\mathcal{S}_{\rm hom}$ is attracting when $0 \leq x_2 < \frac{4}{3}-\frac{\sqrt{10}}{3}$.
            \item For a contracting LRS-II spacetime, $\mathcal{S}_{\rm hom}$ is attracting when $x_2 > 3+\sqrt{7}$.
            \item For a spatially flat ($^{3}R=0$) expanding or contracting LRS-II, there will always be two directions in the phase space along which this invariant submanifold is attracting, and two directions along which it is repelling.
        \end{itemize}
\end{itemize}

\end{enumerate}

Before ending this section, some important discussions are in order:
\begin{itemize}
\item Firstly, it is worth mentioning that for LRS-II, spatially flat solutions do \emph{not} in general represent an invariant submanifold. If we define $x_5 \equiv \frac{3}{2}\frac{^{3}R}{\theta^2}$, then, using the Friedmann equation \eqref{fried_dimless}, one can derive 
\begin{equation}
    \frac{dx_5}{\epsilon d\tau} = \frac{1}{3} x_5 (x_2 + 2x_3 + 4x_3^2) + \frac{3}{2}x_3 (x_1^2 + 2y_1),
\end{equation}
where the second term, which depends explicitly on the shear, prevents $x_5=0$ from being an invariant submanifold. The 3-curvature locally experienced by a comoving observer can change between being positive and negative. Only when the shear vanishes, spatially flat solutions represent an invariant submanifold.

\item For the points at the intersection between the invariant submanifolds $\mathcal{S}_{\Sigma\mathcal{E}}$ and $S_{\mu}$, the evolution asymptotically homogenizes with expansion whereas inhomogeneities grow with contraction. It is easy to find out the behaviour of the radial perturbations for points belonging to the intersection of these two invariant submanifolds, as putting $\{x_2,x_3,x_4\}=\{0,0,0\}$ leads to a linear system for $y_2$ and $z$:
    \begin{subequations}
        \begin{eqnarray}
        && \frac{dy_2}{\epsilon d\tau} = - \frac{1}{3}y_2 \,,\\
        && \frac{dz}{\epsilon d\tau} = - \frac{1}{6}y_2 - \frac{1}{3}z \,,
        \end{eqnarray}
    \end{subequations}
whose solution is $\{y_2(\tau)\,,z(\tau)\}=\{c_1 e^{-\epsilon\tau/3}\,,c_1 e^{-\epsilon\tau/3}\left(c_2 - \frac{1}{6}\epsilon\tau\right)\}$. Using the propagation equations \eqref{prop_dimless}, we can deduce the complete behaviour of the radial perturbations
\begin{equation}
    y_2(\tau) = c_1 e^{-\epsilon\tau/3} = 9y_4(\tau) \,, \quad z(\tau) = c_1 e^{-\epsilon\tau/3}\left(c_2 - \frac{1}{6}\epsilon\tau\right) = y_3(\tau) 
\end{equation}
Evidently, the phase flow is towards $\mathcal{S}_{\rm hom}$ for $\epsilon=+1$ and away from $\mathcal{S}_{\rm hom}$ for $\epsilon=-1$.
\end{itemize}


\section{Finite fixed points}\label{sec: fin_fp}

We can obtain five isolated fixed points and two lines of fixed points for the system \eqref{dynsys_dimless}, which are listed in Table \ref{tab: table_lrs2}.
\begin{table}[h]
\resizebox{\textwidth}{!}{
    \centering
\begin{tabular}{|c|c|c|c|}
\hline
Fixed point & $(x_1,x_2,x_3,x_4,y_2,z)$ & Stability & \begin{tabular}{@{}c@{}} Actual \\ spacetime geometry \end{tabular}
\\
\hline
$\mathcal{L}_1$ & $(\lambda,0,0,0,0,0)$ & \begin{tabular}{@{}c@{}} Attractor for expanding LRS-II \\ Repeller for contracting LRS-II \end{tabular} & \begin{tabular}{@{}c@{}} Minkowski \\ (Milne w.r.t. comoving observer) \end{tabular}
\\
\hline
$\mathcal{L}_2$ & $\left(\lambda,0,-\frac{1}{2},0,0,\frac{1}{2}\lambda\right)$ & Saddle &  Minkowski
\\
\hline
$\mathcal{P}_1$ & $\left(0,0,-1,-\frac{4}{9},0,0\right)$ & \begin{tabular}{@{}c@{}} Repeller for expanding LRS-II \\ Attractor for contracting LRS-II \end{tabular} & \begin{tabular}{@{}c@{}} Homogeneous \\ Spatially flat \\ Vacuum \end{tabular}
\\
\hline
$\mathcal{P}_2$ & $(0,1,0,0,0,0)$ & Saddle & \begin{tabular}{@{}c@{}} Homogeneous \\ Spatially flat \\ Matter dominated \end{tabular}
\\
\hline
$\mathcal{P}_3$ & $\left(0,0,\frac{1}{4},-\frac{1}{16},0,0\right)$ & Saddle & \begin{tabular}{@{}c@{}} Homogeneous \\ Spatially negatively curved \\ Vacuum \end{tabular}
\\
\hline
$\mathcal{P}_4$ & $(0,0,1,0,0,0)$ & Saddle & Minkowski
\\
\hline
\end{tabular}}
    \caption{Finite fixed points that can be obtained from the system \eqref{dynsys_dimless}. The stability can be analyzed by considering the Jacobian eigenvalues. $\mathcal{L}_1$ and $\mathcal{L}_2$ are lines of fixed points, with $\lambda$ denoting the parameter along the line.}
    \label{tab: table_lrs2}
\end{table}

Let us discuss below what kind of solutions these fixed points represent. In interpreting the spacetime geometry corresponding to these fixed points, we must keep in mind that, when a matter is present, we have a preferred choice of the timelike 4-vector $u^{\mu}$ at each spacetime point. The appropriate foliation is the family of hypersurfaces of constant energy density. We can, then, choose the 4-vector $u^{\mu}$ to be normal to this hypersurface of constant energy density. This choice is unique and has an invariant physical meaning. The interpretations of the spacetime geometry for fixed points that correspond to nonvacuum solutions are made with respect to the a comoving observer having this 4-velocity. 

When there is no matter, there is no unique choice of foliation, as any spacelike hypersurface is a hypersurface of uniform energy density equal to zero. Consequently, there is no unique choice for the timelike 4-vector $u^{\mu}$. The interpretations of the spacetime geometry for fixed points that correspond to a vacuum solution are, therefore ambiguous; one has the freedom to switch to a different $u^{\mu}$, which amounts to a different choice of congruence. Therefore, for those vacuum fixed points whose interpretations are not straightforward with respect to a comoving observer (which are, in fact, the case for all the vacuum fixed points except for $\mathcal{L}_1$), we will adopt alternative approaches to find out the \emph{actual} spacetime geometry corresponding to the fixed points. These actual spacetime geometries need not necessarily correspond to the spacetime geometry seen by the comoving observers of the LRS-II geometry.

\begin{enumerate}
    \item $\mathcal{L}_1$: The line of fixed points $\mathcal{L}_1$ lies at the intersections of the invariant submanifold $\mathcal{S}_{\mu}$ and $\mathcal{S}_{\rm hom}$, which implies this is a vacuum and homogeneous solution with respect to the comoving observers. The constraint equation \eqref{fried_dimless} shows that such a fixed point is possible only when the 3-curvature is negative  $^{3}R<0$, which identifies this solution as Milne with respect to the comoving observer. 

    On the other hand, since it is a vacuum case, there is no preferred choice for congruence of timelike geodesics. It is known that one can redefine the coordinates in such a way, i.e. switch to a different congruence of geodesics such that the Milne geometry is recast in a Minkowski form. Another way of arguing that the geometry is effectively Minkowski is that since both matter density (and hence, by virtue of the Einstein field equation, the Ricci tensor) vanishes and electric Weyl tensor vanishes ($x_2 = x_4 = 0 \Rightarrow \mu=\mathcal{E}=0$), the Riemann curvature tensor must vanish. The spacetime geometry is actually Minkowski, although with respect to the LRS-II comoving observers ($u^{\mu}=(1,0,0,0)$), locally it is Milne. 

    Therefore, homogeneous Milne arises as an attractor in an expanding LRS-II and as a repelled in a contracting LRS-II. The attractive nature of the Milne solution in GR was found in Refs.\cite{Coley:2008qd,Wainwright:2009zz} in the study of LTB dynamics. We show that the same conclusion holds for the generic case of LRS-II. 

    \item $\mathcal{L}_2$: The line of fixed points $\mathcal{L}_2$ falls on the invariant submanifold $\mathcal{S}_\mu$. If a comoving observer is going through an evolutionary phase characterized by this fixed point, the matter density locally with respect to the observer vanishes. Moreover, since $y_2=0$, i.e. $\hat{\mu}=0$, the comoving observer also sees no radial perturbation in matter density. Also, the line is characterized by the relation
    \begin{equation}\label{rel1_L2}
        z = \frac{1}{2}x_1 \,\, \Rightarrow \hat{\theta} = \frac{1}{2}\phi\theta.
    \end{equation}
    Also, $x_3 = - \frac{1}{2}$ implies the following relation between the local expansion and shear
    \begin{equation}\label{rel2_L2}
        \theta + 3\Sigma = 0.
    \end{equation}
    $x_3 = - \frac{1}{2}$ also implies, via the constraint equation \eqref{fried_dimless}, that locally the 3-curvature is negative $^{3}R<0$. Since $x_4 = 0$, the electric Weyl tensor locally vanishes. Moreover, from the propagation equation, \eqref{prop_dimless} one can see that $y_2 = 0$ and $x_4 = 0$ also makes $y_4 = 0$, i.e. the comoving observer does not experience any radial perturbation of the electric Weyl tensor. Unlike $\mathcal{L}_1$, it is, however, not straightforward to interpret the spacetime geometry with respect to a comoving LRS-II observer.

    On the other hand, since it is a vacuum case, the Ricci tensor vanishes by virtue of the Einstein field equation. Since the Weyl curvature tensor also vanishes, this definitely means the vanishing of the Riemann curvature tensor. Therefore, the spacetime geometry corresponding to this solution is Minkowski. A Minkowski spacetime satisfies the relations \eqref{rel1_L2},\eqref{rel2_L2}, but is apparently inconsistent with the fact that we are getting $^{3}R<0$. However, since it is a vacuum solution, there is no preferred choice for congruence of timelike geodesics. We can switch to a different congruence, i.e. redefine the coordinates in such a way that the Minkowski geometry is recast in a Milne form, reconciling the fact that we are getting a negative 3-curvature. 

    \item $\mathcal{P}_1$: The isolated fixed point $\mathcal{P}_1$ lies at the intersection of the invariant submanifolds $\mathcal{S}_\phi$, $\mathcal{S}_{\mu}$ and $\mathcal{S}_K$. It represents a vacuum solution. It can be found from the constraint equation \eqref{fried_dimless} that $^{3}R=0$, i.e. the solution is spatially flat. Since $y_2=0$ ($\hat{\mu}=0$) and $z=0$ ($\hat\theta=0$), the comoving observer does not experience any radial perturbation in matter density and the rate of expansion is the same for all the comoving observers. From the propagation equations \eqref{prop_dimless} one can also calculate that $\{y_1,y_3,y_4\}=\{0,0,0\}$, i.e. $\{\hat{\phi},\hat{\Sigma},\hat{\mathcal{E}}\}=\{0,0,0\}$. This implies that the comoving observer experiences no radial perturbation in \emph{any} of the dynamical quantities. All the LRS-II comoving observers see the same spacetime geometry, i.e. the situation is homogeneous. With respect to the LRS-II comoving observers, the spacetime geometry is spatially flat, homogeneous and anisotropic. 

    There is another way of arriving at the conclusion that the spacetime geometry must be homogeneous. This fixed point is a vacuum solution of GR. An extension of Birkhoff's theorem holds for LRS-II spacetimes in GR \cite{Goswami:2011ft,Goswami:2012jf}, stating that the solution possesses an additional symmetry which makes the spacetime either spatially homogeneous with $\phi=0$ or static with $\theta=\Sigma=0$. The condition $\theta=\Sigma=0$ is clearly not satisfied for this fixed point whereas the condition $\phi=0$ is satisfied, suggesting that the spacetime must be homogeneous.

    \item $\mathcal{P}_2$: The isolated fixed point $\mathcal{P}_2$ lies at the intersection of the invariant submanifolds $\mathcal{S}_{\phi}$, $\mathcal{S}_{\rm hom}$ and $\mathcal{S}_K$. From the Friedmann constraint \eqref{fried_dimless}, one can see that the 3-curvature corresponding to this fixed point solution is zero $^{3}R=0$. Therefore, it represents a spatially flat homogeneous solution. A comoving observer describes the spacetime geometry to be a spatially flat and homogeneous. Note that, since this is not a vacuum solution, there is a preferred choice of the congruence of timelike geodesics, namely, those that are along the worldline of the comoving observers. 

    The time-evolution corresponding to this solution can be found out from eq.\eqref{imp_rel}, which when written in terms of the Hubble parameter $H\equiv\frac{\theta}{3}\equiv\frac{\dot a}{a}$, gives 
    \begin{equation}
        \frac{\dot H}{H^2} = - \frac{3}{2}\,.
    \end{equation}
    This corresponds to the time evolution 
    \begin{equation}
        a(t) \sim 
        \begin{cases}
        (+t)^{2/3}\,, \qquad (t>0) \qquad \text{for an expanding universe} \\
        (-t)^{2/3}\,, \qquad (t<0) \qquad \text{for a contracting universe}
        \end{cases}
    \end{equation}
    Since $^{3}R=\Sigma=0$, the original Friedmann equation \eqref{fried} gives, in terms of the Hubble parameter,
    \begin{equation}
        3H^2 = \mu\,.
    \end{equation}
    We, therefore, get the matter-dominated epoch as a saddle point, i.e. intermediate state of LRS-II evolution with dust. 

    \item $\mathcal{P}_3$: The isolated fixed point $\mathcal{P}_3$ lies at the intersection of the invariant submanifolds $\mathcal{S}_{\phi}$, $\mathcal{S}_\mu$ and $\mathcal{S}_K$. From the Friedmann constraint \eqref{fried_dimless}, one can see that the 3-curvature corresponding to this fixed point solution is negative $^{3}R<0$. It is a vacuum solution. Since $y_2=z=0$, a comoving observer does not experience any radial perturbation in matter distribution and expansion rate ($\hat\mu=\hat\theta=0$). From the propagation equations \eqref{prop_dimless} one can also calculate that $\{y_3,y_4\}=\{0,0\}$, i.e. $\{\hat{\Sigma},\hat{\mathcal{E}}\}=\{0,0\}$ but $y_1\neq0$, i.e.$\hat{\phi}\neq0$.

    However, just like the point $\mathcal{P}_1$, the fixed point $\mathcal{P}_3$ is also a vacuum LRS-II solution of GR and therefore possesses an additional symmetry which makes the spacetime either spatially homogeneous or static. Since $\phi=0$ for this fixed point, this suggests that the spacetime must be homogeneous \cite{Goswami:2011ft,Goswami:2012jf}.

    \item $\mathcal{P}_4$: The isolated fixed point $\mathcal{P}_4$ also lies at the intersection of the invariant submanifolds $\mathcal{S}_{\phi}$, $\mathcal{S}_\mu$ and $\mathcal{S}_K$. From \eqref{fried_dimless}, the spatial 3-curvature vanishes $^3R=0$. Therefore, this is a spatially flat vacuum solution. Since this fixed point corresponds to a vacuum situation and the electric part of the Weyl tensor also vanishes ($x_4=0$), the Riemann curvature tensor vanishes, which specifies the actual spacetime geometry to be Minkowski.
\end{enumerate}

Before moving on to the next section about the fixed points at the infinity of the phase space, some discussions are in order regarding the finite fixed points.
\begin{itemize}

    \item In table \ref{tab: table_lrs2}, we have listed the \emph{actual} spacetime geometry corresponding to the fixed points representing vacuum solutions. we emphasize again that these actual spacetime geometries are not always the geometry seen by a comoving observer. With respect to a different congruence, these actual geometries should be apparent. The idea is similar to when one tries to rewrite the static Schwarschild or Minkowski metric with respect to the coordinates attached to a comoving observer, one gets back a non-static metric, but the actual geometry still remains Schwarzschild or Minkowski. In the absence of any matter, the observer always has this freedom. No such ambiguity appears for fixed points that represent nonvacuum solutions.
    
    \item One notices that all the isolated finite fixed points in the phase space of LRS-II in the presence of dust fall at the intersection of the invariant submanifolds $\mathcal{S}_{\phi}$ and $\mathcal{S}_{K}$. The corresponding evolutionary phases are achieved by a comoving observer only when (s)he experiences zero expansion of the 2-sheet along the preferred radial direction and zero Gaussian curvature of the shell. We have given the expressions of $\phi$ and $K$ in terms of the metric functions for the LRS-II metric \eqref{LRS-II} in Eqs.\eqref{phi_lrs2} and \eqref{K_lrs2} respectively. From those expressions, it is not easy to interpret the conditions $\phi=0$ and $K=0$ from the coordinate point of view in its full generality. The interpretation becomes easier when we specialize to a particular metric, e.g. LTB, as we will see in section \ref{sec: ltb}.
    
    \item $\theta\rightarrow-\infty$ represents a \emph{shell-focusing} singularity, when the congruence of comoving geodesics runs into a caustic. On the other hand, $\theta\rightarrow+\infty$ can be interpreted as a \emph{shell-diverging} singularity, when the comoving geodesics are infinitely away from each It is possible for all the finite fixed points listed in the table \ref{tab: table_lrs2} to represent singular situations. Take, for example, the fixed point $\mathcal{P}_2$. It can represent a situation where $\theta$ diverges faster than $\Sigma,\,\mathcal{E}$ and at the same rate as $\sqrt{3\mu}$. This is precisely the big-bang or big-crunch singularity ingrained in an FLRW evolution. In fact, if an expansion normalized dynamical variable $x_i$ vanishes for a fixed point, one possible interpretation of the fixed point can be a singular situation where $\theta$ diverges faster than whatever covariant quantity is there in the numerator of $x_i$.

    \item Perhaps the most remarkable observation is, even if we have started with an inhomogeneous spacetime in the presence of an inhomogeneous matter (dust) distribution, all the finite isolated fixed points that we could find actually represent a homogeneous solution. As we will see in the next section, this is not the case with fixed points at the infinity of the phase space. It is possible for some of those asymptotic fixed points (belonging to the family $\mathcal{B}^{(4)}$; see section \ref{sec: inf_fp}) to represent inhomogeneous geometries. 

\end{itemize}


\section{Fixed points at infinity}\label{sec: inf_fp}

To investigate the global behaviour of the phase space, it is necessary to find fixed points that may reside at the infinity of the phase space. To find the fixed points at the infinity of the phase space, we compactify the 6D phase space spanned by the dynamical variables $x_1,\,x_2,\,x_3,\,x_4,\,y_2,\,z$ using the usual Poincare compactification prescription. The procedure is as follows. First, we define spherical polar coordinates as follows:
\begin{equation}
\begin{aligned}
    x_1 &= r \cos{\alpha}\cos{\beta}\cos{\gamma}\cos{\delta}\cos{\theta}\,,\\
    x_2 &= r \sin{\alpha}\cos{\beta}\cos{\gamma}\cos{\delta}\cos{\theta}\,,\\
    x_3 &= r \sin{\theta}\,,\\
    x_4 &= r \sin{\beta}\cos{\gamma}\cos{\delta}\cos{\theta}\,,\\
    y_2 &= r \sin{\gamma}\cos{\delta}\cos{\theta}\,,\\
    z &= r \cos{\delta}\cos{\theta}\,.
\end{aligned}        
\end{equation}
Now, we have a 6D phase space spanned by the radial variable $r$ and five angular variables $\alpha,\,\beta,\,\gamma,\,\delta,\,\theta$. Next, we compactify the radial direction as
\begin{equation}
    r \rightarrow R = \frac{r}{\sqrt{1+r^2}}\,.
\end{equation}
Now that the phase space is compactified, we go back to Cartesian coordinates again by defining
\begin{equation}
\begin{aligned}
    X_1 &= R \cos{\alpha}\cos{\beta}\cos{\gamma}\cos{\delta}\cos{\theta}\,,\\
    X_2 &= R \sin{\alpha}\cos{\beta}\cos{\gamma}\cos{\delta}\cos{\theta}\,,\\
    X_3 &= R \sin{\theta}\,,\\
    X_4 &= R \sin{\beta}\cos{\gamma}\cos{\delta}\cos{\theta}\,,\\
    Y_2 &= R \sin{\gamma}\cos{\delta}\cos{\theta}\,,\\
    Z &= R \cos{\delta}\cos{\theta}\,.
\end{aligned}        
\end{equation}
The dynamical variables $X_1,\,X_2,\,X_3,\,X_4,\,Y_2,\,Z$ are now compact, limited within the 6-sphere given by
\begin{equation}
    0 \leq X_1^2 + X_2^2 + X_3^2 + X_4^2 + Y_2^2 + Z^2 \leq 1
\end{equation}
The direct transformation between the compact and noncompact dynamical variables are
\begin{equation}\label{dynvar_comp}
    \begin{aligned}
       X_1 = \frac{x_1}{\sqrt{1+r^2}}\,,\quad & X_2 = \frac{x_2}{\sqrt{1+r^2}}\,,\quad X_3 = \frac{x_3}{\sqrt{1+r^2}}\,,\quad X_4 = \frac{x_4}{\sqrt{1+r^2}}\,,
        \\
       & Y_2 = \frac{y_2}{\sqrt{1+r^2}}\,,\quad Z = \frac{z}{\sqrt{1+r^2}}\,.
    \end{aligned}
\end{equation}
and the inverse transformation is
\begin{equation}
    \begin{aligned}
       x_1 = \frac{X_1}{\sqrt{1-R^2}}\,,\quad & x_2 = \frac{X_2}{\sqrt{1-R^2}}\,,\quad x_3 = \frac{X_3}{\sqrt{1-R^2}}\,,\quad x_4 = \frac{X_4}{\sqrt{1-R^2}}\,,
        \\
       & y_2 = \frac{Y_2}{\sqrt{1-R^2}}\,,\quad z = \frac{Z}{\sqrt{1-R^2}}\,.
    \end{aligned}
\end{equation}
The above transformations can be used to write the dynamical system \eqref{dynsys_dimless} in terms of the compact dynamical variables \eqref{dynvar_comp}. The expressions are huge and we do not find it necessary to write their explicit forms here. One encounters a singularity of the dynamical system at the boundary of the phase space $R\rightarrow1$. This can be regularized by redefining the phase space time variable as
\begin{equation}
    \tau\rightarrow\Bar{\tau} : d\bar\tau = \frac{d\tau}{1-R^2}\,.
\end{equation}
In terms of $\bar\tau$, one can get a regular dynamical system in terms of the compact dynamical variables. As we already investigated the finite fixed points in the previous section, in this section we specifically list the asymptotic fixed points, i.e. fixed points with $R=1$.

There are three different families of asymptotic fixed points. They are as follows:
\begin{enumerate}
    \item A pair of isolated fixed points given by $\mathcal{O}_\pm\equiv(X_1,X_2,X_3,X_4,Y_2,Z) = (0,0,\pm1,0,0,0)$, which corresponds to $x_3=\frac{3\Sigma}{2\theta}\rightarrow\pm\infty$. Both these points are saddles.  Both these asymptotic fixed points lie on the invariant submanifolds $\mathcal{S}_\phi$ and $\mathcal{S}_\mu$. They are vacuum solutions. From Eq.\eqref{gauss_curv_dimless}, we find that these fixed points lie on the region $x_K<0$, i.e. the corresponding spacetime geometry, as seen by a comoving observer, must be cylindrically symmetric. Since this is a vacuum case, the Ricci tensor vanishes by virtue of the Einstein field equation. $X_4=0$ imply that the electric Weyl tensor also vanishes. Together, they imply the vanishing of the Riemann tensor. The actual spacetime geometry is therefore Minkowski. A comoving observer, of course, does \emph{not} see the geometry as Minkowski, as evident from the fact that $x_3\equiv\frac{3\Sigma}{2\theta}=\pm\infty$. Nonetheless, with respect to a different timelike congruence, the Minkowski geometry should be apparent. 
    \item A 1-parameter family of fixed points $\mathcal{A}^{(1)}$ given by $X_2=X_4=Y_2=0,\,Z=3X_1,\,10X_1^2+X_3^2=1$. this corresponds to the quantity $(10x_1^2 + x_3^2)$ diverging, $z=3x_1$ and all other variables vanishing. All the fixed points in this family are saddles. This entire family resides on the invariant submanifold $\mathcal{S}_\mu$. Just like $\mathcal{O}_\pm$, the actual spacetime geometry corresponding to all the fixed points within this family is Minkowski.
    \item A 4-parameter family of fixed points $\mathcal{B}^{(4)}$ given by $X_3=0,\, X_1^2+X_2^2+X_4^2+Y_2^2+Z^2=1$ which correspond to $x_3=0$ and the quantity $(x_1^2+x_2^2+x_4^2+y_2^2+z^2)$ diverging. All the fixed points belonging to this family are also saddles. Different fixed points of this family represent different spacetime geometries, but they are all isotropic ($x_3=0$).
\end{enumerate}

The fixed points at infinity of the phase space cannot represent singular situations. Such fixed points are characterized by some expansion normalized dynamical variables $x_i$ going to infinity. Therefore, either $|\theta|\rightarrow0$ or the covariant quantity at the numerator of $x_i$ diverges faster than $\theta$. 


\section{Lema\'itre-Tolman-Bondi (LTB) spacetime}\label{sec: ltb}

The spherically symmetric Lema\'itre-Tolman-Bondi (LTB) spacetime, which belongs to LRS-II class, is of special interest because of its relevance in the study of gravitational collapse \cite{Joshi:1993zg,Joshi:2014gea,Joshi:2007hq,Joshi:2001xi,Joshi:2004tb,Joshi:2002id} and in attempts to explain late-time cosmological observations without invoking any dark energy component \cite{Garfinkle:2006sb,Iguchi:2001sq,Perivolaropoulos:2014lua,Marra:2011ct,Zumalacarregui:2012pq}. Different aspects of LTB spacetime has been studied before using the 1+1+2 formalism \cite{Dunsby:2010ts,Zibin:2008vj}. This section is devoted to describing the LTB spacetime geometry in light of the dynamical system formulation developed in the previous sections.

In terms of 3D spherical polar coordinates LTB metric is written as
\begin{equation}\label{LTB}
    ds^2 = - dt^2 + \frac{R^{\prime 2}(t,r)}{1-\mathcal{K}(r)}dr^2 + R^2(t,r) d\Omega^2.
\end{equation}
For this metric the normalized timelike vector $u^{\mu}$ and the normalized spacelike vector $e^{\mu}$ are, respectively,
\begin{equation}
    u^{\mu} = (1,0,0,0), \qquad e^{\mu} = \left(0,\frac{\sqrt{1-\mathcal{K}(r)}}{R'},0,0\right).
\end{equation}
For this particular choice of vectors, the covariantly defined time and radial derivatives for a scalar quantity $\psi$ become
\begin{equation}\label{cov_devs}
    \dot{\psi} = \partial_t \psi ,\qquad \hat{\psi} = \frac{\sqrt{1-\mathcal{K}(r)}}{R'}\partial_{r}\psi.
\end{equation}
The coordinate dependent expressions of the relevant non-vanishing geometric scalars are as follows \footnote{For the metric functions $\dot{(...)}$ means derivative w.r.t. the time coordinate $t$, not to be confused with the covariant time derivatives of the covariant quantities. $(...)'$ means derivative w.r.t. $r$.}
\begin{subequations}
\begin{eqnarray}
&& \phi \equiv d_{\mu}e^{\mu} = \frac{2\sqrt{1 - \mathcal{K}(r)}}{R}\,, 
\\
&& \theta \equiv D_{\mu}u^{\mu} = \frac{\dot{R}'}{R'} + 2\frac{\dot{R}}{R}\,, 
\\
&& \Sigma \equiv \sigma_{\mu\nu}e^{\mu}e^{\nu}
= D_{\langle\mu}u_{\nu\rangle}e^{\mu}e^{\nu}
= e^{\mu}e^{\nu} \left(h^{\alpha}_{\,\,\,(\mu}h^{\beta}_{\,\,\,\nu)} - \frac{1}{3}h_{\mu\nu}h^{\alpha\beta}\right)D_{\alpha}u_{\beta} 
= \frac{2}{3}\left(\frac{\dot{R}'}{R'} - \frac{\dot{R}}{R}\right)\,, 
\\
&& \mathcal{E} \equiv \frac{1}{6R}\left(\frac{\mathcal{K}'}{R'} - 2\frac{\mathcal{K}}{R}\right) + \frac{1}{3}\left(\frac{\Ddot{R}}{R} - \frac{\Ddot{R'}}{R'}\right) - \frac{1}{3}\frac{\dot R}{R}\left(\frac{\dot R}{R} - \frac{\dot{R'}}{R'}\right)\,, 
\\
&& ^{3}R = 2\frac{(\mathcal{K}R)'}{R^{2}R'}\,,
\\
&& K = \frac{1}{R^2}\,.
\end{eqnarray}
\end{subequations}
The above expressions specifically allow us to interpret the solutions on the invariant submanifold $\mathcal{S}_{\phi}$ and $\mathcal{S}_K$ in a much clearer manner in the context of LTB. Note that one must have $K(r)<1$ in an LTB spacetime. Then, both the conditions $\phi=0$ and $K=0$ represent solutions at spatial infinity $R(t,r)\rightarrow\infty$. Consequently, for an LTB spacetime, the evolutionary phases represented by the isolated finite fixed points $\mathcal{P}_1,\,\mathcal{P}_2,\,\mathcal{P}_3,\,\mathcal{P}_4$ can be achieved by a comoving observer only when the observer is at spatial infinity. 

Note that, since LTB is spherically symmetric, its phase space is bounded by
\begin{equation}
    x_K > 0 \Leftrightarrow x_2 - 9x_4 + \frac{9}{4}x_1^2 - (1 - x_3)^2 > 0\,.
\end{equation}
This implies that the asymptotic fixed point $\mathcal{O}^{\pm}$ is outside the range of the LTB phase space. The other families of asymptotic fixed points $\mathcal{A}^{(1)}$ and $\mathcal{B}^{(4)}$ can be achieved within the LTB phase space.

The existence of the line of attractors $\mathcal{L}_1$ suggests that an expanding LTB cosmology will asymptotically approach Milne. This conclusion is in agreement with those available in the literature \cite{Coley:2008qd,Wainwright:2009zz}.

The original Birkhoff's theorem states that a spherically symmetric vacuum solution in GR is necessarily either a part of the extended Schwarzschild manifold or Minkowski. When the Weyl curvature tensor is non-vanishing, it is necessarily a part of the Schwarzschild spacetime. However, it can be either the static Schwarzschild exterior spacetime ($\theta=\Sigma=0$) or the spatially homogeneous Schwarzschild interior spacetime ($\phi=0$). In light of this, we can conclude that the isolated fixed points $\mathcal{P}_1,\,\mathcal{P}_3$ both represent Schwarzschild interior solutions.

For the sake of completeness, we list in table \ref{tab: table_ltb} the finite and asymptotic fixed points of the LTB phase space. 

\begin{table}[h]
\resizebox{\textwidth}{!}{
    \centering
\begin{tabular}{|c|c|c|c|}
\hline
Fixed point & $(x_1,x_2,x_3,x_4,y_2,z)$ & Stability & \begin{tabular}{@{}c@{}} Actual \\ spacetime geometry \end{tabular}
\\
\hline
$\mathcal{L}_1$ & $(\lambda,0,0,0,0,0)$ & \begin{tabular}{@{}c@{}} Attractor for expanding LTB \\ Repeller for contracting LTB \end{tabular} & \begin{tabular}{@{}c@{}} Minkowski \\ (Milne w.r.t. comoving observer) \end{tabular}
\\
\hline
$\mathcal{L}_2$ & $\left(\lambda,0,-\frac{1}{2},0,0,\frac{1}{2}\lambda\right)$ & Saddle &  Minkowski
\\
\hline
$\mathcal{P}_1$ & $\left(0,0,-1,-\frac{4}{9},0,0\right)$ & \begin{tabular}{@{}c@{}} Repeller for expanding LTB \\ Attractor for contracting LTB \end{tabular} & Scharzschild interior
\\
\hline
$\mathcal{P}_2$ & $(0,1,0,0,0,0)$ & Saddle & \begin{tabular}{@{}c@{}} Spatially flat \\ matter dominated FLRW \end{tabular}
\\
\hline
$\mathcal{P}_3$ & $\left(0,0,\frac{1}{4},-\frac{1}{16},0,0\right)$ & Saddle & Scharzschild interior
\\
\hline
$\mathcal{P}_4$ & $(0,0,1,0,0,0)$ & Saddle & Minkowski
\\
\hline
$\mathcal{A}^{(1)}$ & \begin{tabular}{@{}c@{}} $(10x_1^2 + x_3^2)$ diverging, $z=3x_1$, \\ $x_2 = x_4 = y_2 = 0$ \end{tabular} & Saddle & Minkowski
\\
\hline
$\mathcal{B}^{(4)}$ & \begin{tabular}{@{}c@{}} $(x_1^2 + x_2^2 + x_4^2 +y_2^2 + z^2)$ diverging, \\ $x_3=0$ \end{tabular} & Saddle & Minkowski
\\
\hline
\end{tabular}}
    \caption{Finite and asymptotic fixed points of the LTB phase space.}
    \label{tab: table_ltb}
\end{table}

The conclusions drawn regarding the stability of different invariant submanifolds of the LRS-II phase space, which we listed in section \ref{sec: inv_sub}, were independent of the Gaussian curvature. Therefore they hold in exactly the same form also for LTB. For LTB, the invariant submanifold $\mathcal{S}_{\rm hom}$ contains FLRW solutions and the stability of this submanifold determines the evolution of almost-FLRW LTB geometries. From the generic conclusions that we could draw regarding the stability of $\mathcal{S}_{\rm hom}$ in section \ref{sec: inv_sub}, we can conclude that the homogenization of an almost-FLRW LTB depends crucially on the matter density. An almost-FLRW expanding LTB will isotropize provided the matter density is in the range $0 \le \frac{\mu}{\theta^2} < \frac{4}{9}-\frac{\sqrt{10}}{9}$, whereas an almost-FLRW contracting LTB will isotropize provided $\frac{\mu}{\theta^2} > 1+\frac{\sqrt{7}}{3}$. Qualitatively one can conclude that a large matter density hinders the homogenization of an almost-FLRW expanding LTB whereas assists in the homogenization of an almost-FLRW contracting LTB.


\section{Conclusion}\label{sec: conclusion}

In this work, we present a new dynamical system formulation for LRS-II spacetimes utilizing the 1+1+2 covariant decomposition approach \cite{Clarkson:2002jz,Clarkson:2007yp}. Our approach describes the dynamics of an LRS-II spacetime from the point of view of a comoving observer. A crucial point that we have emphasized throughout the text is that for a comoving observer, the relevant dynamical quantities are not only the covariant geometric and thermodynamic quantities but also their covariantly defined ``radial derivatives'', which we interpret as a specific kind of perturbation that preserves the local rotational symmetry but stops the spacetime from being globally homogeneous. This realization helps us obtain an autonomous dynamical system containing a set of first-order ordinary differential equations involving only time derivatives and a set of purely algebraic constraints. Another nice feature of our formulation is that the dimensionless dynamical variables that span the LRS-II phase space are expansion normalized and directly related to actual covariant geometrical and thermodynamic quantities, making it easier to interpret a fixed point or an invariant submanifold with respect to a comoving observer. Due to the inhomogeneity of the system, or in other words due to the existence of the non-vanishing radial perturbations, not all the comoving observers will go through the same evolutionary phase simultaneously. 

We identify several invariant submanifolds and fixed points in sections \ref{sec: inv_sub}, \ref{sec: fin_fp} and \ref{sec: inf_fp}. We found that homogeneous solutions contain an invariant submanifold whose stability depends only on the quantity $\frac{\mu}{\theta^2}$ (or equivalently $\frac{^{3}R}{\theta^2}$). For LTB, one can qualitatively say that a large matter density hinders the homogenization of an almost-FLRW expanding LTB whereas assists in the homogenization of an almost-FLRW contracting LTB. Interestingly, we find that spatially flat solutions in general do not constitute an invariant submanifold in the LRS-II phase space because of a non-vanishing shear. Finding a solution corresponding to vacuum fixed points becomes ambiguous due to not having a preferred choice for $u^{\mu}$ (i.e. a preferred frame), something which is a precondition for the applicability of the 1+3 formalism. Special care is taken to interpret the solutions corresponding to the vacuum fixed points. We find a spatially flat matter-dominated FLRW solution as a saddle fixed point, i.e. an intermediate epoch of evolution, as one might expect in the presence of nonrelativistic matter. We also recovered a future attractor which a comoving observer would identify with Milne, consistent with earlier results that expanding LTB solutions without the cosmological constant tends to Milne \cite{Coley:2008qd,Wainwright:2009zz}.

Let us finish with an outlook of possible further research in this line. The logical next step is to generalize the present analysis including a cosmological constant term (e.g. as in the $\Lambda$LTB model \cite{Sundell:2016uqj}) and/or a non-vanishing pressure. In this work, we got a decoupled set of evolution and propagation equations because we chose to confine our attention to perfect fluids. Also, for perfect fluids the worldlines of comoving observers are geodesics, since the acceleration vector $a^{\mu}$ vanishes. However, neither the aforementioned decoupling nor the vanishing of acceleration is a necessary condition to obtain an autonomous dynamical system in our approach; they just merely make the calculations simpler. Even if one considers a non-perfect fluid one could still proceed following our approach and construct an autonomous dynamical system that describes the dynamics from the point of view of a comoving observer. It is interesting to explore a formal comparison between the 1+1+2-based dynamical system formulation that we developed here with the dynamical system based on the orthonormal frame approach \cite{vanElst:2001xm,Uggla:2003fp,Lim:2003ut} and Sussman's approach in terms of quasi-local variables \cite{Sussman:2007ku,Sussman:2010zp}; whether or not there exists a one-to-one correspondence between fixed points and invariant submanifolds. It might be interesting to revisit the question whether the cosmic no-hair theorem (an inflating spacetime isotropizes asymptotically) holds in the presence of inhomogeneity, a question which has been addressed previously using the orthonormal frame formalism \cite{Lim:2003ut}. Apart from cosmology, we expect that this formulation should also find vast application in the study of gravitational collapse  \cite{Joshi:2000fk,Joshi:2011zm,Malafarina:2016let}. In this context, it is interesting to see if we can somehow incorporate the information regarding the apparent horizon in the phase space of a collapsing LTB and subsequently address the cosmic censorship conjecture \footnote{For a study of cosmic censorship in the 1+1+2 language, see \cite{Hamid:2014kza,Hamid:2016rnf}.}. All these interesting lines of research are reserved as potential future projects.

\section*{Acknowledgement} 

SC acknowledges funding support from the NSRF via the Program Management Unit for Human Resources and Institutional Development, Research and Innovation (Thailand) [grant number B39G670016]. PKSD acknowledges the First Rand Bank (SA) for financial support. 


\begin{appendix}
    
\section{An alternate approach to phase space analysis for $^3R>0$}\label{sec: comp_Rpos}

For non-negative spatial curvature $^{3}R>0$, it is possible to formulate a dynamical system with an alternative choice of dynamical variables. The formulation we present here has the merit that for some homogeneous spacetimes it can lead to a compact phase space, allowing for a global phase space analysis \cite{Solomons:2001ef}. For LRS-II, as we will see below, they do not lead to a compact phase space. Nonetheless, it is still interesting to explore what this alternative formulation gives us and how it compares with the usual formulation in terms of the expansion normalized dynamical variables that we have considered in the main body of the paper. 

Inspired by the compact phase space formalism used in \cite{Solomons:2001ef}, we define the following quantities
\begin{eqnarray}
&& D \equiv \sqrt{\frac{\theta^2}{9} + \frac{^3R}{6}} = \sqrt{\frac{\mu}{3} + \frac{\Sigma^2}{4}}\,, \\
&& U_1 \equiv \frac{\phi}{D}, \quad U_2 \equiv \frac{\mu}{3D^2}, \quad U_3 \equiv \frac{\Sigma}{2D}, \quad U_4 \equiv \frac{\mathcal{E}}{D^2}, \quad U_5 \equiv \frac{\theta}{3D}, \quad U_6 \equiv \sqrt{\frac{^3R}{6}}\frac{1}{D}\,.\label{dynvar_def_1}
\end{eqnarray}
The Friedmann equation implies following two constraints 
\begin{equation}
    U_5^2 + U_6^2 = U_2 + U_3^2 = 1,
\end{equation}
using which we choose to eliminate $U_2$ and $U_6$. At the tangent space of the observer, two sets of dynamical quantities are relevant: the original covariant quantities as well as their covariant radial variations, given by the hat derivatives. Therefore, we also define the following variables
\begin{equation}
    V_1 \equiv \frac{\hat\phi}{D^2}, \quad V_2 \equiv \frac{\hat\mu}{3D^3}, \quad V_3 \equiv \frac{\hat\Sigma}{2D^2}, \quad V_4 \equiv \frac{\hat{\mathcal{E}}}{D^3}, \quad V_5 \equiv \frac{\hat\theta}{3D^2}\,.\label{dynvar_def_2}
\end{equation}
Together, Eqs.\eqref{dynvar_def_1} and \eqref{dynvar_def_2} provide the full set of dynamical variables with respect to which we can write a closed set of autonomous equations describing the LRS-II dynamics. 


Clearly, except for $U_3$ and $U_5$, no other dimensions in the phase space is compactified; $-1<U_3,U_5<1$. However, $U_5$ being bounded makes it easier to characterize the singularity; a big bang ($\theta\rightarrow+\infty$) corresponds to $U_5\rightarrow1$ and a big crunch ($\theta\rightarrow-\infty$) corresponds to $U_5\rightarrow-1$. It is to be kept in mind that although singularities are represented by specific points in the phase space, in the local $u$-$e$ plane they are represented by curves (singularity curve).

The propagation equations \eqref{prop} imply the following constraint equations
\begin{subequations}\label{prop_dimless_1}
    \begin{eqnarray}
        V_1 &=& -\frac{1}{2}U_1^2 + 2(U_5 + 2U_3)(U_5 - U_3) - 2U_2 - U_4\,,\\
        V_3 - V_5 &=& -\frac{3}{2}U_1 U_3\,,\\
        V_4 - V_2 &=& -\frac{3}{2}U_1 U_4\,.
    \end{eqnarray}
\end{subequations}
using which we choose to eliminate $V_1,\,V_3,\,V_4$.

The quantity $D$ satisfies the following relations
\begin{equation}
    \frac{\dot{D}}{D^2} = - \frac{3}{2}U_5 - \frac{1}{2}U_3(2U_3^2 + U_5 U_3 + U_4), \label{evoln_D}
\end{equation}
We define the following time variable on the phase space $\tau_c$ as
\begin{equation}
    d\tau_c = D dt\,.
\end{equation}
We are now in a position to write the dynamical system in terms of the dynamical variables defined in Eqs.\eqref{dynvar_def_1} and \eqref{dynvar_def_2}. Using the relation \eqref{evoln_D}, we can write the system as
\begin{subequations}\label{dynsys_dimless_1}
    \begin{eqnarray}
    && \frac{dU_1}{d\tau_c} = U_1 \left[\frac{1}{2}U_5 + U_3 + \frac{1}{2}U_3(2U_3^2 + U_5 U_3 + U_4)\right] \,,\\
    && \frac{dU_3}{d\tau_c} = - \frac{1}{2}U_4 - U_3(U_3 + 2U_5) + \frac{1}{2}U_3^2(2U_3^2 + U_5 U_3 + U_4) \,,\\
    && \frac{dU_4}{d\tau_c} = -3U_3(1 - U_4 - U_3^2) + U_3 X_4(2U_3^2 + U_5 U_3 + U_4) \,,\\
    && \frac{dU_5}{d\tau_c} = - \frac{1}{2}(1 - U_5^2) - \frac{3}{2}U_3^2 + \frac{1}{2}U_5 U_3(2U_3^2 + U_5 U_3 + U_4) \,,\\
    && \frac{dV_2}{d\tau_c} = \frac{1}{2}V_2(U_5 - 4U_3) - 3V_5(1 - U_3^2) + \frac{3}{2}U_3 V_2 (2U_3^2 + U_5 U_3 + U_4) \,,\\
    && \frac{dV_5}{d\tau_c} = 6 U_1 U_3^2 - \frac{1}{2}V_2 - 6U_3 V_5 + U_3 V_5 (2U_3^2 + U_5 U_3 + U_4) \,.\\
    \end{eqnarray}
    \end{subequations}
The above system presents only two fixed points: $P_{\pm}\equiv\{U_1,U_3,U_4,U_5,V_2,V_5\}=\{0,0,0,\pm1,0,0\}$, which represents big-bang and big-crunch singularities. Both of them are saddle fixed points. Both these fixed points have $U_2=1$ and $U_6=0$, i.e. matter dominates the spacetime dynamics on the approach to the singularities whereas the contributions from the 3-curvature asymptotically vanishes. $P_{\pm}$ resides on the invariant submanifold $\mathcal{S}_{\rm hom}$, which now is given by $\{U_3,U_4,V_2,V_5\}=\{0,0,0,0\}$, implying that these are isotropic singularities. 

The mapping between the sets of dynamical variables $\{U_i,V_i\}$ and $\{x_i,y_i\}$ are as follows
\begin{equation}\label{mapping}
    \begin{aligned}
    & x_1=\frac{U_1}{3U_5}, \quad x_2=\frac{U_2}{U_5^2}, \quad x_3=\frac{U_3}{U_5}, \quad x_4\equiv\frac{U_4}{9U_5^2},\\
    & y_1=\frac{V_1}{9U_5^2}, \quad y_2=\frac{V_2}{3U_5^3}, \quad y_3=\frac{V_3}{3U_5^2}, \quad y_4=\frac{V_4}{27U_5^3},\\
    & z=\frac{V_5}{3U_5^2}.
    \end{aligned}
\end{equation}
Substituting back the coordinates $\{U_i,V_i\}$'s of $P_\pm$ in Eq.\eqref{mapping}, we get the coordinates $\{x_1,x_2,x_3,x_4,y_2,z\}=\{0,1,0,0,0,0\}$, which corresponds to the spatially flat matter-dominated homogeneous fixed point $\mathcal{P}_2$ tn Tab.\ref{tab: table_lrs2}. In the LTB phase space this point represents the spatially flat matter dominated FLRW fixed point (Tab.\ref{tab: table_ltb}). The result is consistent since both $P_{\pm}$ and $\mathcal{P}_2$ are saddles. Therefore, the fixed points $P_{\pm}$ represent the big bang and big crunch singularities already ingrained in the matter dominated homogeneous solutions. The singular nature becomes apparent in the phase space formulation presented in this section. However, this phase space formulation fails to give us any of the other fixed points.

\end{appendix}

\bibliography{refs}
\bibliographystyle{unsrt}

\end{document}